\documentclass{article}
\pdfoutput=1

\usepackage[utf8]{inputenc}
\usepackage{amsmath,amsthm}
\usepackage{amsfonts}
\usepackage{amssymb}
\usepackage{epic,eepic,latexsym,color}
\usepackage{ifthen,graphics,epsfig}
\usepackage[section]{algorithm}
\usepackage{algorithmic}
\usepackage{subfigure}
\usepackage{xspace}
\usepackage{threeparttable}
\usepackage{fullpage}

\newcommand{\Amitabh}[1]{**\textbf{Amitabh: \color{blue}[[[#1]]]}**} 
\newcommand{\Armando}[1]{**\textbf{Armando:} \emph{#1}**} 
\newcommand{\dd}[1]{{\textbf{**DD: \color{red}[[[#1]]]}**}}
\renewcommand{\Armando}[1]{}
\renewcommand{\Amitabh}[1]{}
\renewcommand{\dd}[1]{}

\newcommand{\Atemp}[1]{**\textbf{A Temp: \color{blue}[[[#1]]]}**} 
\renewcommand{\Atemp}[1]{}

\newcounter{algocf}
\addtocounter{algocf}{1}
\newcounter{linecounter}
\newcommand{\linenumbering}{\ifthenelse{\value{linecounter}<10}
{(0\arabic{linecounter})}{(\arabic{linecounter})}}

\renewcommand{\thelinecounter}{\ifnum \value{linecounter} > 
9\else 0\fi \arabic{linecounter}}

\def\beginsmall#1{\vspace{-\parskip}\begin{#1}\itemsep-\parskip}
\def\endsmall#1{\end{#1}\vspace{-\parskip}}

\newtheorem{theorem}{Theorem}[section]
\newtheorem{lemma}[theorem]{Lemma}

\newenvironment{oldthm}[1]{\par\noindent{\bf Theorem #1:} \em \noindent}{\par}
\newenvironment{oldlem}[1]{\par\noindent{\bf Lemma #1:}
  \em \noindent}{\par}
\newenvironment{oldcor}[1]{\par\noindent{\bf Corollary #1:} \em \noindent}{\par}
\newenvironment{oldpro}[1]{\par\noindent{\bf Proposition #1:} \em \noindent}{\par}

\newcommand{\ethm}{\end{theorem}}
\newcommand{\commentout}[1]{}
\newcommand{\othm}[1]{\begin{oldthm}{\ref{#1}}}
\newcommand{\eothm}{\end{oldthm} \medskip}
\newcommand{\olem}[1]{\begin{oldlem}{\ref{#1}}}
\newcommand{\eolem}{\end{oldlem} \medskip}
\newcommand{\ocor}[1]{\begin{oldcor}{\ref{#1}}}
\newcommand{\eocor}{\end{oldcor} \medskip}
\newcommand{\opro}[1]{\begin{oldpro}{\ref{#1}}}
\newcommand{\eopro}{\end{oldpro} \medskip}

\newcommand{\ie}{\emph{i.e.,\ }}

\newcommand{\portsym}[1]{{<}#1{>}}

\newcommand{\RT}{\mathrm{RT}}
\newcommand{\ID}{\mathrm{ID}}

\newcommand{\hparent}{\mathrm{hparent}}

\newcommand{\children}{\mathrm{children}}

\newcommand{\FT}{\mathrm{FT}}

\newcommand{\degree}{\mathrm{deg}}

\newcommand{\numchildren}{\mathrm{numchildren}}
\newcommand{\heir}{\mathrm{Heir}}
\newcommand{\parent}{\mathrm{Parent}}
\newcommand{\helper}{\mathrm{helper}}
\newcommand{\real}{\mathrm{Real}}

\newcommand{\will}{\mathrm{Will}}
\newcommand{\leafwill}{\mathrm{LeafWill}}
\newcommand{\willportion}{\mathrm{Willportion}}
\newcommand{\willportions}{\mathrm{Willportions}}
\newcommand{\leafwillportion}{\mathrm{LeafWillportion}}
\newcommand{\leafwillportions}{\mathrm{LeafWillportions}}
\newcommand{\leafheir}{\mathrm{LeafHeir}}

\newcommand{\hchildren}{\mathrm{hchildren}}

\newcommand{\diam}{\mathrm{Dia}}

\newcommand{\execute}{\mathrm{Execute}}

\newcommand{\bypass}{\mathrm{Bypass}}
\newcommand{\Empty}{\emph{EMPTY}}

\newcommand{\fieldnumchildren}{\textbf{numchildren}}
\newcommand{\fieldmaxportnum}{\textbf{maxportnumber}}

\newcommand{\CFT}{\mathrm{CompactFT}}
\newcommand{\emphCFT}{\emph{CompactFT}}

\newcommand{\tzft}{\mathrm{CompactFTZ}}
\newcommand{\tz}{\mathrm{TZ}}
\newcommand{\emphtzft}{\emph{CompactFTZ}}

\newcommand{\lrindex}{\mathrm{\ell}}

\newcommand{\tzheavy}{\mathrm{tzheavy}}
\newcommand{\tzlight}{\mathrm{tzlight}}
\newcommand{\emphtzheavy}{\emph{tzheavy}}
\newcommand{\emphtzlight}{\emph{tzlight}}

\newcommand{\BrNodeReplace}{\mathrm{BrNodeReplace}}
\newcommand{\BrLeafLost}{\mathrm{BrLeafLost}}
\newcommand{\ptwillconnection}{\mathrm{PtWillConnection}}
\newcommand{\ptnewleafwill}{\mathrm{PtNewLeafWill}}

\newboolean{short}
\setboolean{short}{true}

\newcommand{\onlyShort}[1]{\ifthenelse{\boolean{short}}{#1}{}}
\newcommand{\onlyLong}[1]{\ifthenelse{\boolean{short}}{}{#1}}

\newcommand{\squishlist}{
 \begin{list}{$\bullet$}
  { \setlength{\itemsep}{0pt}
     \setlength{\parsep}{3pt}
     \setlength{\topsep}{3pt}
     \setlength{\partopsep}{0pt}
     \setlength{\leftmargin}{1.5em}
     \setlength{\labelwidth}{1em}
     \setlength{\labelsep}{0.5em} } }

\newcommand{\squishlisttwo}{
 \begin{list}{$\bullet$}
  { \setlength{\itemsep}{0pt}
     \setlength{\parsep}{0pt}
    \setlength{\topsep}{0pt}
    \setlength{\partopsep}{0pt}
    \setlength{\leftmargin}{2em}
    \setlength{\labelwidth}{1.5em}
    \setlength{\labelsep}{0.5em} } }

\newcommand{\squishend}{
  \end{list}  }

\DeclareGraphicsExtensions{.pdf}

\title{
Compact Routing Messages in Self-Healing Trees
}

\author{
Armando Casta\~neda\\ Instituto de Matem\'aticas, UNAM
\and Danny Dolev\\The Hebrew University of Jerusalem
\and Amitabh Trehan \thanks{Corresponding Author. Telephone: +447466670830 (Cell), email: a.trehan@qub.ac.uk} \\School of Electronics, Electrical Engineering and Computer Science,\\ Queen's University Belfast, UK
}

\begin{document}

\maketitle

\begin{abstract}

Existing compact routing schemes, e.g., Thorup and Zwick [SPAA 2001] and Chechik [PODC 2013], often have no means to tolerate failures, once the system has been setup and started. This paper presents, to our knowledge, the first self-healing compact routing scheme. Besides, our schemes are developed for low memory nodes, i.e., nodes need only $O(\log^2 n)$ memory, and are thus, compact schemes.

We introduce two algorithms of independent interest: The first is $\emphCFT$, a novel compact version (using only $O(\log n)$ local memory) of the self-healing algorithm Forgiving Tree of Hayes et al. [PODC 2008]. The second algorithm ($\emphtzft$) combines CompactFT with Thorup-Zwick's tree-based compact routing scheme [SPAA 2001] to produce a fully compact self-healing routing scheme. In the self-healing model, the adversary deletes nodes one at a time with the affected nodes self-healing locally by adding few edges. $\CFT$ recovers from each attack in only $O(1)$ time and $\Delta$ messages, with only +3 degree increase and  $O(log \Delta)$ graph diameter increase, over any sequence of deletions ($\Delta$ is the initial maximum  degree). 

Additionally, $\tzft$  guarantees delivery of a packet sent from sender $s$ as long as the receiver $t$ has not been deleted, with only an additional  $O(y \log \Delta)$ latency, where  $y$ is the number of nodes that have been deleted on the path between $s$ and $t$. If $t$ has been deleted, $s$ gets informed and the packet removed from the network.  

\end{abstract}




\section{Introduction}

\Amitabh{Routing in real networks uses ARP (Address Resolution Protocol) that takes IP addresses and gives MAC addresses? (at least in data centres). Therefore, there could be a justification for getting a routing address used just for routing, which will be, say, different for the address needed to make a healing edge}

\Amitabh{develop Self-healing Compact routing which will enable low-memory devices (such as sensor fitted `things') to route messages even when some of them may fail (e.g. when your smart watch, smart shirt and google glass walk into a lift that has no wi-fi reception). }

Efficient routing is becoming critical in current networks, and more so in future networks.  
Routing protocols have been the focus of intensive research over the years.  
Routing is based on information carried by the traveling packets and data structures that are maintained at the intermediate nodes.
The efficiency parameters change from time to time, as the network use develops and new bottlenecks are identified.  
It is clear that the size of the network eliminates the ability to use any centralized decisions, and we are close to giving up on maintaining long distance routing decisions.  
We are a few years before a full scale deployment of IOT (Internet of Things) that will introduce billions of very weak devices that need to be routed.
The size of the network and the dynamic structure that will evolve will force focusing on local decisions. The weakness of future devices and the size of the network will push for the use of protocols that do not require maintaining huge routing tables.


Santoro and Khatib~\cite{santoro1985labelling}, Peleg and Upfal~\cite{PelegU88}, and Cowen~\cite{Cowen:1999} 
 pioneered the concept of compact routing that requires only a minimal storage at each node.  Moreover, the use of such routing protocols imposes only a constant factor increase on the length of the routing.  Several papers followed up with some improvements on the schemes (cf. Thorup and Zwick~\cite{ThorupZ01}, Fraigniaud and Gavoille~\cite{RoutingFraigniaudG01}, and Chechik~\cite{Chechik13}).
These efficient routing schemes remain stable as long as there are no changes to the network.  

\Atemp{Added definition of compact below}
The target of the current paper is to introduce an efficient \emph{compact} scheme that combines compact routing with the ability to correct the local data structure stored at each node in a response to the change.  Throughout this paper, when we say compact, we imply schemes that use $o(n)$ local memory (in our case, we actually only use $O(\log^2 n)$ local memory) per node. Our new scheme has similar cost as previous compact routing schemes. We will focus on node failures, since that is more challenging to handle.

%
%

\Armando{What I understand is that CompactFT requires a particular setting on the port numbering in the tree,
so we do not start with any arbitrary graph. One thing we should say, maybe in the final discussion, is that TZ 
can be combined with FT which assumes nothing particular, and in that case the network can be initially an
arbitrary connected graph.}

 Our algorithms work in the \emph{bounded memory self-healing mode}l (Section~\ref{sec: model}).  We assume that the network is initially a connected graph over $n$ nodes. 
All nodes are involved in a preprocessing stage in which the nodes identify edges to be included in building a spanning tree over the network and construct their local data structures. 
The adversary repeatedly attacks the network. The adversary knows the network topology and the algorithms, and has the ability to delete arbitrary nodes from the network.
To enforce a bound on the rate of changes, it is assumed the adversary is constrained in that it deletes one node at a time, and between two consecutive deletions, nodes in the neighbourhood of the deleted node can exchange messages with their immediate neighbours and can also request for additional edges to be added between themselves.


 Our self-healing algorithm $\CFT$ ensures recovery from each attack in only a constant time and $\Delta$ messages, while,  over any sequence of deletions, taking only constant additive degree increase (of 3) and keeping the diameter as $O(D \log \Delta)$, where  $D$ is the diameter of the initial graph and $\Delta$ the maximum degree of the initial spanning tree built on the graph. Moreover, $\CFT$ needs only $O(\log n)$ local memory (where $n$ is the number of nodes originally in the network). Theorem~\ref{thm: cftmain} states the results formally.

$\tzft$, our compact routing algorithm, is based on the compact routing scheme on trees by Thorup and Zwick~\cite{ThorupZ01}, and ensures routing between any pair of existing nodes in our self-healing tree without loss of any message packet whose target is still  connected. Moreover, the source will be informed if the receiver is lost,  and if both the sender and receiver have been lost,  the message will be discarded from the system within at most twice of the routing time. 
Our algorithm guarantees that after any sequence of deletions, a packet from $s$ to $t$ is routed through a path of length
$O(d(s,t) \log \Delta)$, where $d(s,t)$ is the distance between $s$ and $t$,  
 and $\Delta$ is the
   maximum degree of any node in the initial tree. Though $\tzft$  uses only $\log n$ local memory, the routing labels (and, hence, the messages) are of $O(\log^2 n)$ size, so nodes may need $O(\log^2 n$) memory to locally process the messages. Theorem~\ref{theo-1} states the results formally.

\Atemp{Add challenges in designing the algorithms; answer why $\tzft$ is not obvious given $\CFT$}

A few modifications are needed to  \cite{ThorupZ01} to  achieve our compact fault-tolerant scheme. We have to ensure that the packets are routed despite the deletions, subsequent self-healing, and, thus, loss and obsolescence of some information, i.e., we would like to continue without updating outdated routing  tables and labels.  This is partly achieved by using a post-order DFS labelling that allows the self-healing nodes to do routing using just simple binary search in the affected areas (\cite{ThorupZ01} uses pre-order DFS  labels). Other modifications to the algorithm and labels modify the routing according to the self-healing state of the nodes, ensure packets are not lost while self-healing and allow undelivered packets to dead targets to be returned to alive senders.




\paragraph{Related Work}

\begin{table}[th!]


{\footnotesize
\begin{threeparttable}[b]
\begin{tabular}{|l|c|cc|ccc|}
\hline
Algorithm & &\multicolumn{2}{|c|}{Over Complete Run} &  \multicolumn{3}{|c|}{Per Healing Phase}\\ 
 \hline
 &Local & Diameter & Degree &  Parallel & Msg  &    \# Msges \\
  & Memory& (Orig: $D$)& (Orig: $d$)&  Repair Time &  Size &   \\
 \hline
 Forgiving Tree~\cite{HayesPODC08} & $O(n)$  & $D \log\Delta $\tnote{$\dagger$} & $d + 3$ &   $O(1)$ & $O(\log n)$  & $O(1)$  \\
   \hline
CompactFT(this paper) & $O(\log n)$  & $D \log\Delta $\tnote{$\dagger$} & $d + 3$ &  $O(1)$ & $O(\log n)$  & $O(\delta)$\tnote{$\ddagger$}  \\
  \hline
 \end{tabular}
 \begin{tablenotes} 
 \item[$\dagger$] $\Delta$: Highest degree of network.
  \item[$\ddagger$] $\delta$: Highest degree of a node involved in repair (at the most $\Delta$).
\end{tablenotes} 
\caption{\small Comparison of CompactFT with Forgiving Tree}
\label{tab: AlgoCompare}
\end{threeparttable}
} 
\end{table}

$\CFT$ uses ideas from the \emph{Forgiving Tree}~\cite{HayesPODC08} ($\FT$, in short) approach in order to improve compact routing.  
The main improvement of $\CFT$ is that no node uses more than $O(\log n)$ local memory
and thus, $\CFT$ is compact. 
$\CFT$ achieves the same bounds and healing invariants as $\FT$, however, taking slightly more messages (at most $O(\Delta)$ messages as opposed to $O(1)$ in $\FT$) in certain rounds. Table~\ref{tab: AlgoCompare} compares both algorithms.  



Several papers have studied the routing problem in arbitrary networks
(e.g. \cite{AwerbuchBLP90, AwerbuchGPV90, Cowen01, Chechik13}) and with the help of
geographic information (e.g. \cite{BoseMSU01, FraserKU08, KranakisSU99}), but without failures.
These papers are interested in the trade-off between the size of the routing tables and the stretch of the scheme:
the worst case ratio between the length of the path obtained by the routing scheme and the 
length of the shortest path between  the source and the destination. 
 Here we are mainly interested in preserving compactness under the presence of failures.

 An interesting line of research deals with labelling schemes. \cite{KormanP-IC-Label07} presents labelling schemes for weighted dynamic trees where node weights can change. However, it does not deal with node deletions nor does it claim to deal with routing.
In~\cite{KormanPR-TCS-Label04}, Korman et al. present a compact distributed labelling scheme in the dynamic tree model: 
 (1) the network is a tree, (2) nodes are deleted/added one at a time, (3) the root is never deleted and (4) only leaves can be added/deleted. In a labelling scheme, each node has a  label and from every two labels, the distance between the corresponding nodes can be easily computed. The fault-tolerant labelling scheme is obtained by modifying any static scheme. 
Using the previous, 
they get fault-tolerant (in the same model)  compact versions of the compact tree routing schemes of~\cite{FraigniaudG-STACS-Space02,RoutingFraigniaudG01,ThorupZ01}.
These schemes have a multiplicative overhead factor of $\Omega(\log n)$ on the label sizes of the static schemes. In~\cite{Korman-DC-CompactLabel07}, Korman improves the results in~\cite{KormanPR-TCS-Label04}, presenting a labelling scheme in the same model that allows to compute any function on pairs of vertices with smaller labels, at the cost, in some cases, of communication. 
Our work differs from the previous in the sense that though we use a spanning tree of the network, our network can be arbitrary and any node can be deleted by the adversary.

%
%
%
%
%


There have been numerous papers that discuss strategies for adding additional  
capacity and rerouting in anticipation of failures \cite{ChechikLPR12, Chechik13InfComp, CourcelleT07, 
doverspike94capacity, FengH01, frisanco97capacity,
iraschko98capacity, murakami97comparative, caenegem97capacity, VounckxDLP94, xiong99restore}. 
In each of these solutions, the network is fixed and either redundant information is precomputed or
routing tables are updated when a failure is detected. In contrast, our algorithm
runs in a dynamic setting where edges are added to the network as node
failures occur, maintaining connectivity and  preserving compactness at all time. Our bounded memory self-healing model builds
upon the model introduced in ~\cite{HayesPODC08}. A variety of self-healing topology maintenance algorithms have been devised in 
the self-healing models~\cite{Pandurangan2014-DEX, PanduranganXhealT14, FG-DCJournal2012, Amitabh-2010-PhdThesis, SaiaTrehanIPDPS08}.
Our paper moves in the direction of self-healing computation/routing along with topology which is attempted in other papers e.g. \cite{SaadS-SHComputation-14} 
(though in a different model). Finally, dynamic network topoplogy and fault tolerance are core concerns of distributed computing~\cite{Attiya-WelchBook,Lynchbook}
and various models, e.g.,~ \cite{Kuhn-DistComputation-STOC10}, and topology maintenance and  Self-* algorithms abound~\cite{Berns09DissectingSelf-*,Djikstra74SelfStabilizing,DolevBookSelfStabilization, Dolev09EmpireofColoniesSelf-stabilizing,KormanKMPODC11,Kuhn2005Self-Repairing,Poor-SelfHealQueue2003,Ghosh07Self-healingSystemsSurvey,  BeauquierBK-JThCS11, ElkinSpanners07, BaswanaSenSpanner-Icalp03, KuttenPorat-DynSPanningTree-DC99}.




%




\Amitabh{Mostly follows Forgiving Tree paper: Modify: Add 'bounded local memory'}

\subsection{Bounded Memory  Self-healing Model}

\label{sec: model}

Let $G = G_0$ be an arbitrary connected (for simplicity) graph on $n$ nodes, which represent processors in a distributed network.  Each node has a unique $ID$. Initially, each node only knows its neighbors in $G_0$, and is unaware of the structure of the rest of  $G_0$. 
%

We allow a certain amount of pre-processing to be done before the adversary is allowed to delete nodes.  
%
In the pre-processing stage nodes  exchange messages with their neighbors and setup data structures as required. 


The adversarial process can be described as deleting some node $v_t$ from $G_{t-1}$, forming $H_t$.
All neighbors of $v_t$ are informed of the deletion. 
In the healing stage nodes of $H_t$ communicate (concurrently)
with their immediate neighbors. 
Nodes  may insert edges
joining them to  other nodes they know about, as desired. 
Nodes may also drop edges from previous phases if no longer required.
The resulting graph at the end of this phase is $G_t$. Nodes are not explicitly informed when the healing stage ends. 
We make no synchronicity assumption except that all messages after the deletion of a node, i.e., in a healing phase, are safely received and processed before the adversary deletes the  next node.

The objective is minimizing the following  ``complexity'' measures (excluding the preprocessing stage):
\beginsmall{itemize}
\vspace{-0.5em}
\item{\bf Degree increase:} $\max_{t<n} \max_{v} (\degree(v,G_t) - \degree(v,G_0))$
\item {\bf Diameter stretch:} $\max_{t<n} \diam(G_t) / \diam(G_0)$
\item{\bf Communication:} The maximum number of bits sent by a single node in each recovery phase
\item{\bf Recovery time:} The maximum total time for a recovery phase,
assuming it takes a message no more than $1$ time unit to traverse any edge
\item{\bf Local Memory:} The amount of memory a single node needs to maintain to run the algorithm.
\endsmall{itemize}

\subsection{Compact Routing Model}

The algorithm is allowed a pre-processing phase, e.g., to run a distributed DFS on the graph. Each message has a label which may contain the node ID and other information derived from the pre-processing phase. Every node stores some local information for routing. The routing algorithm does not change the original port assignment at any node (however, node deletions may force the self-healing algorithm to do simple port re-assignments). We are interested in minimizing the sizes of the label and the local information at each node. 
\section{The Algorithms: High Level}
\label{sec: highlevelalgos}

\begin{figure}[t!]
\centering
\includegraphics[scale=0.45]{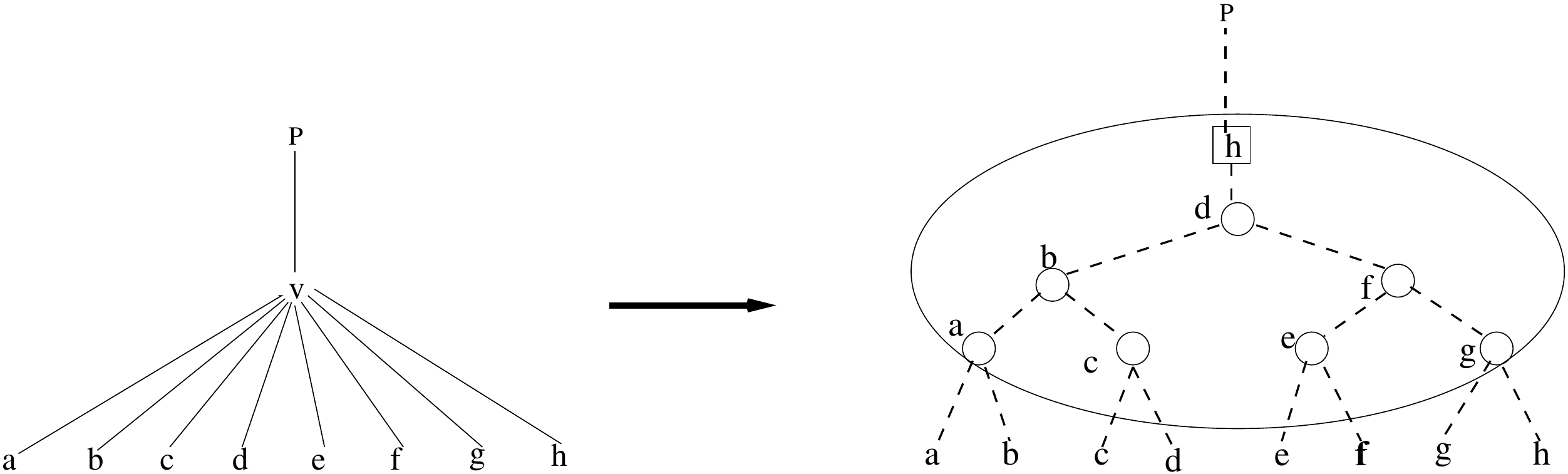}
\caption{ Deleted node $v$ replaced by Reconstruction Tree ($\RT(v)$). Nodes in oval are virtual helper nodes. The circles are regular helper nodes and the rectangle is `heir' helper node. The `Will' of $v$ is $\RT(v)$, i.e., the structure that replaces the deleted $v$. }
 \label{fig: RT}
 \vspace{-1em}
\end{figure}

As stated, $\CFT$ (Algorithm~\ref{algo: cftbrief}) is an adaptation of $\FT$~\cite{HayesPODC08} for low memory ($O(\log n)$) nodes). $\tzft$ (Algorithm~\ref{algo: compactftz}) then conducts reliable routing over $\CFT$. At a high level, the following happens:

\squishlisttwo{
\item \emph{Preprocessing:} A BFS spanning tree $T_0$ of graph $G_0$ is derived followed by DFS traversal and labelling and careful setup of $\CFT$ and $\tzft$ fields (Tables~\ref{tab: cftdata} and ~\ref{tab: tzfields}). For $\CFT$, every node sets up and distributes a $\will$ (Section~\ref{sec: cft}), which is the blueprint of edges and virtual (helper) nodes to be constructed upon the node's deletion. 
\item After each deletion, the repair maintains the spanning tree of helper and real (original) nodes, i.e., the $i^{th}$ deletion (say, of node $v_i$) and subsequent repair yields tree $T_i$.
The helper nodes are simulated by the real nodes and only a real node can be deleted. The two main cases  are as follows:\\
 - \emph{non-leaf deletion, i.e., $v_i$ is not a leaf in $T_{i-1}$} (Section~\ref{subsec: nonleafdeletion}): The neighbours of $v_i$ `execute' $v_i$'s will leading  $v_i$ to be replaced by a \emph{reconstruction tree} ($\RT(v_i)$) which is a balanced binary search tree (\emph{BBST}) 
 having $v_i$'s neighbours as the leaves and $\helper$ nodes  as the internal nodes  simulated by $v_i$'s neighbours (Figure~\ref{fig: RT}).\\ 
 - \emph{leaf deletion, i.e., $v_i$ is a leaf in $T_{i-1}$} (Section~\ref{subsec: leafdeletion}): This case is more complicated in the low memory setting since a node (in particular, $v_i$'s parent $p$) cannot store the list of its children nor recompute its $\will$. If $p$ was dead, $v_i$'s siblings essentially deletes a redundant helper node while maintaining the structure. If $p$ is alive, no new edges are made but $p$ orchestrates a distributed  update of its will while being oblivious of the identity of its children. Thus, when $p$ is eventually deleted, the right structure gets put in place.
\item \emph{Routing:} Independent of the self-healing, a node $s$ may send a message to node $r$ (along with $s$'s own label) using the $\tzft$ protocol. The label on the message (for $r$) along with the local routing fields at a real node tells the next node, say $w$, on the path. If, however, $w$ had been deleted earlier, there could be  a helper node on that port which is part of $\RT(w)$. Now, the message would be routed using the fact that the $\RT(w)$ is a BBST and make it to the right node at the end of $\RT(w)$. The message eventually reaches $r$, but if $r$ is dead, the message is `returned to sender' using $s$'s label.
}
\squishend

\section{CompactFT: Detailed Description}
\label{sec: cft}

\begin{table}[th!]
\begin{tabular}{|p{0.32\textwidth}|p{0.6\textwidth}|}
\hline 
\textbf{Current fields}& Fields having  information about a node's current neighbors\\ \cline{2-2}
  \texttt{parent(v)}& Parent of $v$ \\
  \texttt{parentport(v)} & Port at which $parent(v)$ is attached \\
  \texttt{numchildren(v) }& Number of children of $v$\\
  \texttt{maxportnumber(v)} & Maximum port number used by $v$\\
 \texttt{heir(v),$\portsym{heir(v)}$} & The heir of $v$ and its port\\
\hline
\textbf{Helper fields} & Fields for a helper node $v$ may be simulating. (Empty if none)\\ \cline{2-2}
 \texttt{hparent(v)}& Parent of the helper node  $v$ simulates \\
 \texttt{hchildren(v)}& Children of helper node $v$  simulates.\\
\hline
\textbf{Reconstruction fields / WillPortion /LeafWillPortion} & Fields used by $v$ to reconstruct its connections when its neighbor is deleted.\\ \cline{2-2}
 \texttt{nextparent(v)}& Node which will be  next $\parent(v)$ \\
 \texttt{nexthparent(v)}& Node which will be next $\hparent(v)$\\
 \texttt{nexthchildren(v)}& Node(s) that will be  next $\hchildren(v)$\\
\hline
\textbf{Flags}& Boolean fields telling node's  status.\\ \cline{2-2}
\texttt{hashelper(v)}& True if $v$ is simulating a \emph{helper} node\\
\hline
\end{tabular}
\caption{The fields maintained by a node $v$ for Compact FT.  Each reference to a sibling is tagged  with the port number at which it is attached to parent (not shown above for clarity) e.g. $nextparent(v)$ is $nextparent(v), \portsym{nextparent(v)} $. }
\label{tab: cftdata}
\end{table}

\begin{table}[t!]
\begin{tabular}{|p{0.32\textwidth}|p{0.6\textwidth}|}
\hline 
\textbf{Message}& Description\\ 
\hline
$\BrLeafLost$ $(\portsym{x})$ & Node $v$ broadcasts, informing that \\
& the leaf node at $v$'s port $\portsym{x}$ has been deleted.\\
  \hline
          $\BrNodeReplace$ $((x,\portsym{x}),(h,\portsym{x}))$&  Node $v$ broadcasts, asking receivers  to replace 
    (in their $willportions$)  at $v$'s port $\portsym{x}$,  node $x$ with node $h$. \\
     \hline
    $\ptwillconnection$ $((y,\portsym{y}),(z,\portsym{z}))$  & Node $v$ asks receivers (in their $v.\willportion$) 
   to make an edge between node $y$ and node $z$\\
     \hline
      $\ptnewleafwill$ $((y,\portsym{y}),$ $(z,\portsym{z}),W(y))$  & Node $v$ informs node $z$ that it is the new $\leafheir(y)$ 
      and gives it $W(y)$ (= $\leafwill(y)$). \\
 \hline
\end{tabular}
\caption{Messages used by $CompactFT$ (sent by a node $v$).}
\label{tab: CFTmessages}
\end{table}


In this section, we describe $\CFT$.
$\CFT$ maintains connectivity in an initially connected network with only a total constant degree increase per node and $O(\log \Delta)$ factor diameter increase over any sequence of attacks while using only $O(\log n)$ local memory (where $n$ is the number of nodes originally in the network).
  The formal theorem statement is given in Theorem~\ref{thm: cftmain}. 

 \Armando{Somewhere we should list all our assumptions.} \Amitabh{Attempted above. Does the previous line need to be in the Model section?}
\Amitabh{Actually, I think we don't need second assumption for CompactFT! We may need it for Compact Routing, I think. Removing.}
\Armando{Maybe the intro should has a table summarizing the improvements of compactFT?} 
\Amitabh{Added a table in Intro.}

\begin{figure}[t!]
\centering
\includegraphics[scale=0.75]{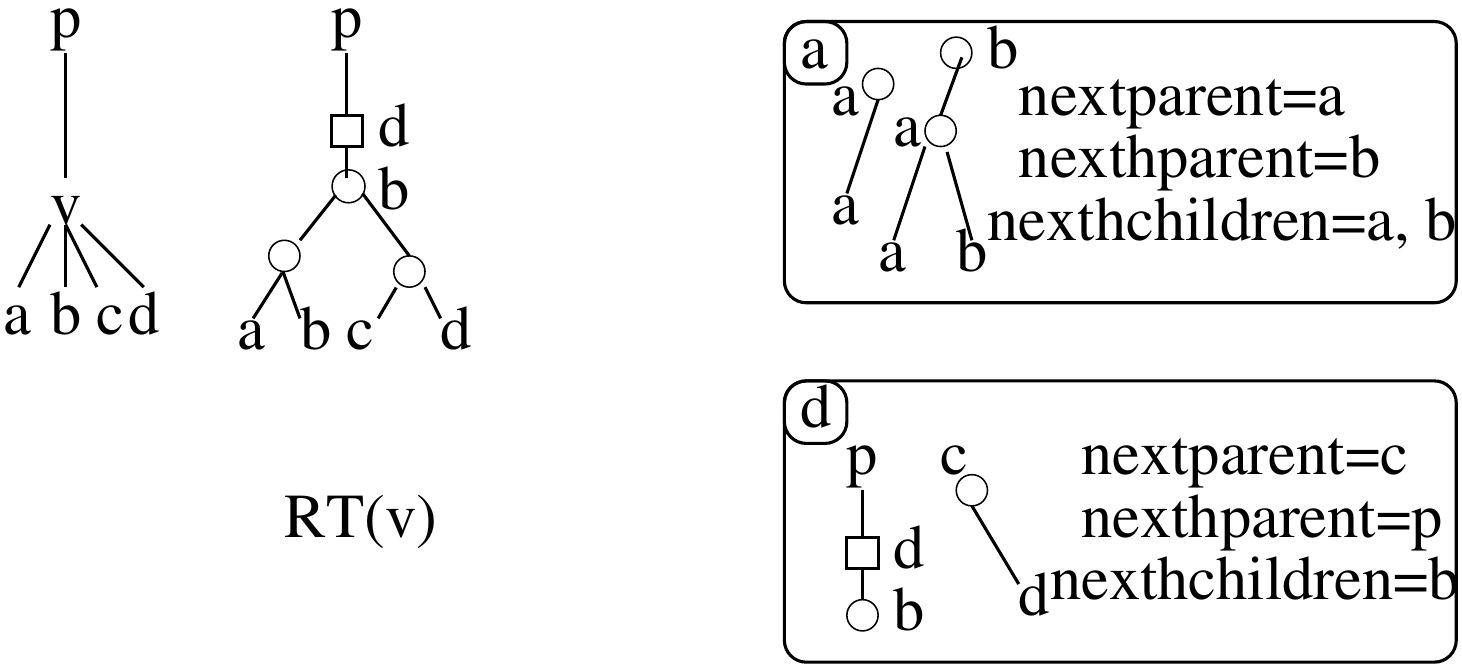}
\caption{The network of a node $v$, $v$'s $\will = \RT(v)$ and $\willportions$ for its  children $a$ and $d$.}
 \label{fig: will}
\end{figure}

As stated, in $\CFT$,
a deleted node is replaced by a 
$\RT$  formed by (virtual)  helper nodes simulated by its children (siblings in case of a leaf deletion) (Figure~\ref{fig: RT}). This healing is carried out by a mechanism of wills:\\
\textbf{Will Mechanism:} 
A $\will(v)$ is the set of actions, i.e., the subgraph to be constructed on the deletion of node $v$. When $v$ is a non-leaf node, this is essentially the  encoding of the structure $\RT(v)$ and is distributed among $v$'s children. Each part of a node's $\will$ stored with another node is called a $\willportion$. We denote $\willportion(p,v)$ to be the part of $\will(p)$ that involves $v$, i.e., the relevant subgraph, and  is stored by node $v$.  When $v$ is a leaf node, however, $\will(v)$ differs in not being a $\RT(v)$ and is stored with siblings of $v$. For clarity, we call this kind of $\will$ a $\leafwill$, the $\willportions$ as $\leafwillportions$ and a leaf node's heir as $\leafheir$. Thus, a $\will$ of a node is distributed among the node's neighbours such that the union of those $\willportions$ makes the whole $\will$. Note that a $\willportion$ (or $\leafwillportion$) is of only constant size (in number of node $ID$s). Figure~\ref{fig: will} shows the $\will$ of a node $v$ and the corresponding $\willportions$.

The fields used by a node for executing  $\CFT$ are given in Table~\ref{tab: cftdata}. Unlike $\FT$, the node cannot have either the list of its children or its own $\RT$ (since these can be as large as $O(n)$).
Rather, a node $v$ will store the number of its children (\emph{numchildren(v)}), highest port number in use  (\emph{maxportnumber}) and  will store every node reference in the form $(node ID, port)$, e.g., in $\willportion(p,v)$, a reference to a node $x$ will be stored as $(x,\portsym{x})$, where $\portsym{x}$ is the port of $p$ at which $x$ is connected.

\begin{algorithm}[h!]
\caption{CompactFT(Graph $G$): High level view}
\label{algo: cftbrief}
\begin{algorithmic}[1]
\STATE \label{algoline: init}\emph{Preprocessing and INIT:} A rooted BFS spanning tree $T(V,E')$ of $G(V,E)$ is computed. For every node $v$, its Will (non-leaf or leaf as appropriate) is computed.  Every node $x$ in a Will is labeled as $(x,\portsym{x})$, where $x$ is $x$'s ID and $\portsym{x}$ is $x$'s parent's  port number at which $x$ is connected (if it exists). Each node only has a $\willportion$ and/or $\leafwillportions$ ($O(\log n)$ sized portion of parent's or sibling's Will, respectively).
\WHILE {true}
\IF{a vertex $x$ is deleted}
\IF[Fix non leaf deletion.]  {$x$ was not a leaf (i.e., had any children)}\STATE $x$'s children \emph{execute} $x$'s Will using $x$'s $\willportions$ they have; $\heir(x)$ takes over $x$'s Will/duties. \label{algoline: fixnonleafstart} 
\STATE All Affected Wills (i.e.  neighbours of $x$ and of $\helper(x)$) are updated by simple update of relevant $\willportions$.  \label{algoline: fixnonleafend}
\ELSE[Fix leaf deletion.]  
\STATE \label{algoline: fixleafstart}
 Let node $p$ (if it exists) be node $x$'s parent \COMMENT{If $p$ does not exist, $x$ was the only node in the network, so nothing to do}
\IF[Update Wills by simulating the deletion of $p$ and $x$]  {$p$ is real/alive} 
\IF{$x$ was $p$'s only child}
 \STATE $p$ computes its $\leafheir$ and $\leafwill$ and forwards it. \COMMENT{$p$ has become a leaf}
 \ELSE
\STATE $p$ informs all children about $x$'s deletion.
\STATE $p$'s children update $p$'s $\willportions$ using $x$'s $\leafwillportions$.
\STATE Children issue updates to $p$'s $\willportions$ and other $\leafwillportions$ via $p$.
\STATE $p$ forwards updates via broadcast or point-to-point messages, as required.
\STATE $p$'s neighbours receiving these messages update their data structures.
\ENDIF
\ELSE[$p$ had already been deleted earlier.] \label{algoline: fixnonleafvp}
\STATE Let $y$ be $x$'s $\leafheir$.
\STATE $y$ \emph{executes} $x$'s $\will$.
\STATE \label{algoline: fixnonleafvpend} Affected nodes update their and their neighbour's $\willportions$.
\ENDIF
\ENDIF 
\IF{$x$ was node $z$'s $\leafheir$} \label{algoline: fixleafheirstart}
\STATE $z$ sets a new neighbour as $\leafheir$ following a simple rule.
\ENDIF
\ENDIF 
\ENDWHILE
\end{algorithmic}
\end{algorithm}

Algorithm~\ref{algo: cftbrief} gives a high level overview of $\CFT$. Algorithms~\ref{algo: cft} through \ref{algo: updatewp} (in the appendix) give a more detailed technical description of the algorithm. Also, Table~\ref{tab: CFTmessages} describes some of the special messages used by $\CFT$.
 $\CFT$  begins with a preprocessing phase (Algo~\ref{algo: cftbrief} line~\ref{algoline: init}) in which 
  a rooted BFS spanning tree of the network from an arbitrary node is computed. The algorithm will then maintain this tree in a self-healing manner.
   Each node sets up the $\CFT$ data structures including its $\will$.
   We do not count the resources involved in the preprocessing but note that at the end of that phase all the $\CFT$ data structures are contained within the $O(\log n)$ memory of a node.
  As stated, the basic operation is to replace a (non-leaf) node by a $\RT$. 
   A leaf deletion, however, leads to a reduction in the number of nodes in the system and the structure is then maintained by a combination of a `short circuiting' operation and a $\helper$ node reassignment (this is also encoded in the leaf node's $\leafwill$ and is discussed later). An essential invariant of $\CFT$ is that \emph{a real node simulates at most one $\helper$ node} and since each $\helper$ node is a node of a binary tree, the degree increase of any node is restricted to at most 3. 
 Similarly, since $\RT$s are balanced binary trees,  
  distances and, hence, the diameter of the $\CFT$, blows up by at most a $\log \Delta$ factor, where $\Delta$ is the degree of the highest node (ref: Theorem~\ref{thm: cftmain}). In the following description, we sometimes refer to a node $v$ as $\real(v)$ if it is real, or $\helper(v)$ if it is a helper node, or by just $v$ if it is obvious from the context.
  
 
 \subsection{Deletion of a Non-Leaf Node:}
\label{subsec: nonleafdeletion}
Assume that a node $x$ is deleted. If $x$ was not a leaf node (Algorithm~\ref{algo: fixnode}, Algorithm~\ref{algo: cftbrief} lines ~\ref{algoline: fixnonleafstart} - \ref{algoline: fixnonleafend}), it's neighbours simply execute $x$'s $\will$. 
One of $x$'s children (by default the rightmost child) is a special child called the $\heir$ (say, $h$) and it takes over any virtual node (i.e., $\helper(x)$) that $x$ may have been simulating, otherwise it is the one that connects the rest of the $\RT$ to the parent of $x$ (say, $p$). This past action may lead to changes in the Wills of other live nodes. In particular, $p$ will have to tell its children to replace $x$ by $h$ in $p$'s $\will$. Due to the limited memory, $p$ does not know the identity of $x$. However, when $h$ contacts $p$, it will inform $p$ that $x$ has been deleted and $p$ will broadcast a message $\BrNodeReplace((x,<x>),(h,<x>))$ asking all neighbours to replace $x$ by $h$ in their $\willportions$ at the same port (Algorithm~\ref{algo: fixnode} and Table~\ref{tab: CFTmessages}).
%

\subsection{Deletion of a Leaf Node:} 
\label{subsec: leafdeletion}

If the deleted node $x$ was, in fact, a leaf node, the situation is more involved. There are two cases to consider:
whether the parent $p$ of $x$ is a helper node  (implying the original parent had been deleted earlier) or a real node. 
The second case, though trivial in $\FT$ (with $O(n)$ memory) is challenging in $\CFT$. Before we discuss the cases, we introduce the `short-circuiting' operation used during leaf deletion:

\beginsmall{description}
\item[bypass(x):] (from ~\cite{HayesPODC08}) \emph{Precondition}: $|\hchildren(x)| = 1$, \ie the helper node has a single child.\\
 \emph{Operation}:
\emph{Delete} $\helper(x)$, i.e., $\parent$ of  $\helper(x)$ and child of $\helper(x)$  remove their edges with $\helper(x)$ and make a
new edge between themselves. \\
 $\hparent(x) \leftarrow \Empty$; $\hchildren(x) \leftarrow \Empty$.
\endsmall{description}

\begin{enumerate}
\item \label{ldcasehelper}  \emph{Parent $p$ of $x$ is a $\helper$ node }

If $p$ is a $\helper$ node, this implies that the original parent of $x$ (in $G_0$) had been deleted at some stage and $x$ has exactly one $\helper$ node in one of the $\RT$s in the tree above. Since $x$ has been deleted, $p$ has only one child now and $\bypass(p)$ can now be executed. 
There are two further cases:

\begin{figure}[t!]
\centering
\subfigure[$\helper(v)$ is the parent of $v$.]{\label{sfig: ldc1} \includegraphics[scale=0.4]{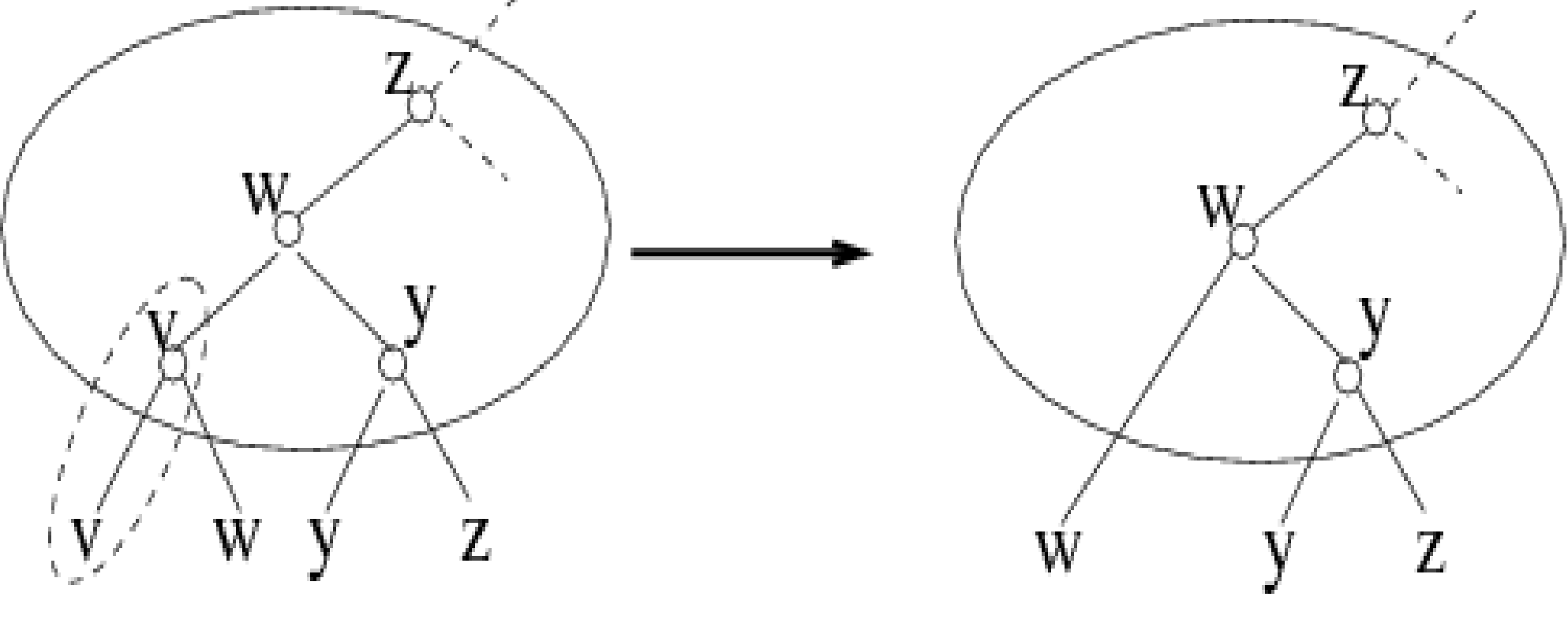}}\hspace{0.3in}
\subfigure[$\helper(w)$ is not the parent of $w$.]{\label{sfig: ldc2} \includegraphics[scale=0.4]{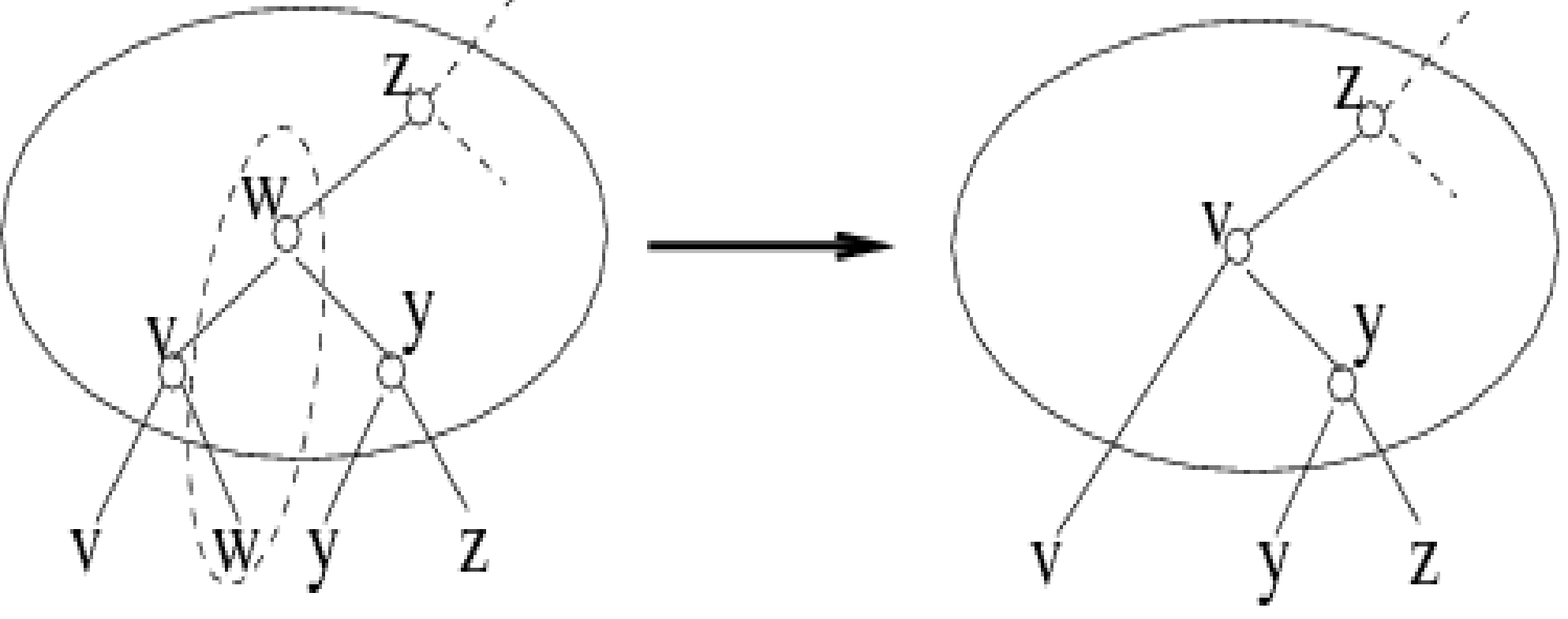}}\\
\vspace{-1em}
\caption{ Deletion of a leaf node whose parent is a $\helper$ node: two cases.}
\label{fig: leafdels}
\end{figure} 

\beginsmall{enumerate}

\item \label{ldhelperself}  \emph{$\helper(x)$ is parent of $x$ (Figure~\ref{sfig: ldc1})}: 
 In this case, the only thing that needs to be done is $\bypass(x)$ since this bypasses the deleted nodes and restores connectivity. However, the issue is that $\helper(x)$ has already been deleted, so how is $\bypass(x)$ to be executed?   For this, we use the mechanism of a $\leafwill$. Assume $\helper(x)$ had two children, $x$ and $y$.
When $x$ sets up its $\leafwill$ (which consists  only  of $\bypass(x)$), it designates $y$ as its $\leafheir$ 
and sends it its $\leafwill$.
In Figure~\ref{sfig: ldc1}, the $\leafheir$ of node $v$ is  $w$ and the $\leafwill(v)$ consists of the operation $\bypass(v)$.

\item  \label{ldhelpernotself}  \emph{$\helper(x)$ is not the parent of $x$ (Figure~\ref{sfig: ldc2})}:
   Let $p$ be the parent of the deleted node $x$.
    Since $p$ now has only one child left, it will have to be short-circuited by $\bypass(p)$. However, the node $\helper(x)$ has also been lost. Therefore, if we don't fix that, we will disconnect the neighbours of $\helper(x)$. However, since $p$ has been bypassed, $\real(p)$ is not simulating a helper node anymore and, thus, $\real(p)$ will take over the slot of $\helper(x)$ by making edges between its ex-neighbours. In this case, $x$ simply designates $p$ as its $\leafheir$ and leaves $\leafwill(x)$ (which is of only $O(\log n)$ size) with $p$. In Figure~\ref{sfig: ldc2}, node $w$ is deleted, its parent and $\leafheir$ is $\helper(v)$ and, thus, when $w$ is deleted, following $\leafwill(w)$,  $\bypass(v)$ is executed and $v$ takes over $\helper(w)$.

\endsmall{enumerate}

The only situation left to be discussed is when $x$ was a $\leafheir$ of another node. In this case, the algorithm follows the rules apparent from the cases before. Let $v$ be the node that had $x$ as its $\leafheir$. Assume that after healing, $p$ is the parent of $\real(v)$ and assume for now that $p$ is a $\helper$ node (the real node case is discussed later). Then, if $p$ is $\helper(v)$, $v$ makes the other child of $p$ (i.e., $v$'s sibling) as $v$'s $\leafheir$, otherwise $v$ sets $p$ as its $\leafheir$ and hands its Will over to the $\leafheir$.

\item \label{ldcasereal} \emph{Parent $p$ of $x$ is a real node} 

 This case is trivial in $\FT$ as all that $p$ needs to do is remove $x$ from the list of its children ($\children(p)$ in $\FT$), recompute its Will and distribute it to all its children. However, in $\CFT$, $p$ cannot store the list of its children and thus, update its Will. Therefore, we have to find a way for the Will to be updated in a distributed manner while still taking only a constant number of rounds. This is accomplished by using the facts that the $\willportions$ are already distributed pieces of $p$'s $\will$ and each leaf deletion affects only a constant number of other nodes allowing us  to update the $\willportions$ locally using Algorithm~\ref{algo: updatewp}. Notice that since $p$ is real, nodes cannot really execute $x$'s $\will$ as in case~\ref{ldcasehelper}. However,  $\will(p)$ is essentially the blueprint of $\RT(p)$. Hence, what $p$ and its neighbours do is  execute  $\will(x)$ on $\will(p)$: this has the effect of updating $\will(x)$ to its correct state and when ultimately $p$ is deleted, the right structure is in place. 
 
 This `simulation' is done in the following manner: $p$ detects the failure of $x$ and informs all its neighbours by a $\BrLeafLost(\portsym{x})$ message (Table~\ref{tab: CFTmessages}). The node that is $\leafheir(x)$, say $v$, will now simulate execution of $\leafwill(x)$. As discussed in Case~\ref{ldcasehelper}, a $\leafwill$ has two parts: a $\bypass$ operation and a possible helper node takeover by another node. Suppose the $\bypass$ operation is supposed to make an edge between nodes $a$ and $b$. Node $v$ simulates this by asking $p$ to send a $\ptwillconnection((a,\portsym{a}),(b,\portsym{b}))$ message to its ports $\portsym{a}$ and $\portsym{b}$. This has the effect of node $a$ and $b$ making the appropriate edge in their $\willportion(p)$. Similarly, for the node take over of $\helper(x)$, $v$ asks $p$ to send $\ptwillconnection$ messages to make edges (in $\willportions$) between the node taking over and the previous neighbours of  $\helper(x)$ in $\will(p)$. 
 
  Another case  is when $x$ was the $\leafheir$ of another node, say $w$. Since $\leafheir(x)$ has already done the healing, the $\willportions$ are now updated and it is easy for $w$  to find another $\leafheir$. This is straightforward as per our previous discussion. The new $\leafheir$ will either be $\real(w)$'s $\parent$ or (if $\parent(w) = \helper(w)$) $\parent(w)$'s other child. Notice this information is already present in $\willportion(p,w)$. The new $\leafwill(w)$ is also straightforward to calculate. As stated earlier, every $\leafwill$ has a $\bypass$ and/or a node takeover operation. All the nodes involved are neighbours of $w$ in $\willportion(p,w)$. Therefore, this information is also available with $w$ enabling  it to reconstruct its new $\leafwill$ which it then sends to the new $\leafheir$ via $p$ using the $\ptnewleafwill()$ message (Table~\ref{tab: CFTmessages}). Finally, there is a special case:

\begin{itemize}
\item \label{ldcasereal} \emph{ $x$ was the only child of  (real) parent $p$}:

Finally, there is also the possibility of node $x$ being the only child of its parent $p$ in which case $p$ will become a leaf itself on $x$'s deletion. Node  $p$ can only be a $\real$ node (a $\helper$ node cannot have one child) and since $x$ does not have any sibling, $x$ will not have any $\leafheir$ or $\leafwill$ (rather, these fields will be set to NULL). Thus, when $x$ will be deleted, there will be no new edges added. However, $p$ will detect that it has become a leaf node and using $p$'s parent's $\willportion$, it will designate a new $\leafheir$, compute a new $\leafwill$ (as discussed previously) and send it to its $\leafheir$ by messages (if $p$'s parent is $\real$) or directly.
\end{itemize}

\end{enumerate}

Theorem~\ref{thm: cftmain} (proof deferred to Section~\ref{sec: cft-apdx})  summarises the properties of $\CFT$.
\begin{theorem}
\label{thm: cftmain}
The $\CFT$ has the following properties:
\beginsmall{enumerate}
\item \label{thm: cftdeg} $\CFT$ increases  degree of any vertex by only 3.
\item \label{thm: cftdiam} $\CFT$ always has diameter $O(D \log \Delta)$, where $D$ is the diameter  and $\Delta$ the maximum degree of the initial graph.
\item \label{thm: cftmem} Each node in $\CFT$ uses only $O(\log n)$ local memory  for the algorithm.
\item \label{thm: cftcosts} The latency per deletion is $O(1)$ and the number of messages sent per node per deletion is $O(\Delta)$; each message contains $O(1)$ node $\ID$s and thus $O(\log n)$ bits.
\endsmall{enumerate}
\end{theorem}


\section{A Compact Self Healing Routing Scheme}
\label{sec: tzft}

In this section, we present a fault tolerant / self-healing routing scheme.
First, we present a  variant of the compact routing scheme on
trees of Thorup and Zwick~\cite{ThorupZ01} (which we refer to as $\tz$ in what follows), and then we make
this algorithm fault-tolerant in the  self-healing model using $\CFT$ (Section~\ref{sec: cft}).
We call the resulting scheme $\tzft$.

\subsection{Compact Routing on Trees}
\label{subsec: tzftstatic}

We present a  variant of $\tz$ that mainly
 differs 
  in the order of DFS labelling of nodes.
The local fields of each node are changed accordingly. 
This variant allows us to route even in the presence of 
 adversarial deletions on nodes when combined with $\CFT$ 
  (Lemma~\ref{lemma-bst}).

Let $T$ be a tree rooted at a node $r$.
Consider a constant $b \geq 2$.
The \emph{weight} $s_v$ of a node $v$ is the number of its descendants, including $v$.
A child $u$ of $v$ is \emph{heavy} if $s_u \geq s_v / b$,
and \emph{light} otherwise. 
Hence, $v$ has at most $b-1$ heavy children.
By definition, $r$ is heavy.
The \emph{light routing index} $\lrindex_v$ of $v$ is the number of light nodes on
the path from $r$ to $v$, including $v$ if it is light. We label a heavy node as \emphtzheavy\ and a light node as \emphtzlight.
Note that here we are describing the scheme for the static case where the tree does not change over time.
However, it is easy to extend this to the dynamic case (Section~\ref{subsec: tzftdynamic}) by initially setting up the data structure in exactly the
same way as the static case during preprocessing. Later on the classification into heavy and light type remains as it was set initially and need not be updated. 

We first enumerate the nodes of $T$ in \emph{DFS post-order manner}, 
with the heavy nodes traversed before the light nodes.
For each node $v$, we let $v$ itself denote this number. 
This numbering gives the $\ID$s of nodes
(in the original scheme, the nodes are labelled in a pre-order manner and 
the light nodes are visited first). For ease of description, by abuse of notation, in the description and algorithm, we refer interchangeably to both the node itself and its $\ID$ as $v$.

Note that each node has an $\ID$ that is larger than the $\ID$ of any of its descendants.
Moreover, given a node and two of its children $u$ and $v$ with $u < v$,
the IDs in the subtree rooted at $u$ are strictly smaller than the $\ID$s
in the subtree rooted at $v$. With such a labelling, routing can be easily performed:
if a node $u$ receives a message for a node $v$, it checks if $v$ belongs to
the interval of $ID$s of its descendants; if so, it forwards the message to 
its appropriate children, otherwise it forwards the message to its parent.
 Using the notion of $\tzlight$ and $\tzheavy$ nodes, one can achieve a compact scheme. The local fields for a node are given in Table~\ref{tab: tzfields}. Note that each node $v$ locally stores $O(b \log n)$ bits.
The \emph{label $L(v)$} of $v$ is defined as follows: an array with the port numbers reaching 
the light nodes in the path from $r$ to $v$.
The definition of $\tzlight$ nodes implies that 
the size of $L(v)$ is $O(\log^2 n)$,
hence the size of the header $(v, L(v))$ of a packet to $v$ is $O(\log^2 n)$.
The scheme TZ  is described in  Algorithm~\ref{algo: tz}.\\


\begin{table}[t!]
\begin{tabular}{|p{0.3\textwidth}|p{0.6\textwidth}|}
\hline
$v$ & DFS number (post-order)\\ 
 \hline
$d_v$ & smallest descendent of $v$ \\ 
 & (in the original scheme, this is $f_v$, the largest descendant of $v$)\\ 
 \hline
$c_v$ & smallest descendent of first $\tzlight$ child of $v$, if it exists; otherwise $v + 1$  \\  
 & (in the original scheme, this is $h_v$, the first $\tzheavy$ child of $v$) \\ 
 \hline
\multicolumn{2}{|l|}{$H_v$:  array with $b+1$ elements}\\ 
 \hline
$H_v[0]$ & number of $ \tzheavy$ children of $v$ \\ 
$H_v[1, \hdots, H_v[0]]$ & $\tzheavy$ children of $v$\\ 
 \hline
\multicolumn{2}{|l|}{$P_v$: array with $b+1$ elements}\\ 
 \hline
$P_v[0]$ & port number of the edge from $v$ to its parent. \\ 
 $P_v[1, \hdots, H_v[0]]$ &  port numbers from $v$ to its $\tzheavy$ children\\ 
 \hline
$\lrindex$ & light routing index of $v$\\ 
\hline
\end{tabular}
\caption{\textbf{Local fields of a node $v$:} Locally, each node $v$ stores the above information}
\label{tab: tzfields}
\end{table}

\begin{algorithm}[h!]
{\bf operation}  ${\sf TZ}_{v} (w, L(w))$:  
\begin{algorithmic}[1]
\IF{$v = w$}
\STATE The message reached its destination
\ELSIF{ $w \notin [d_v, v]$}
\STATE Forward to the parent through port $P_v[0]$  
\ELSIF{$w \in [c_v,v]$}
\STATE Forward to a $\tzlight$ node through port $L(w)[\ell_v]$
\ELSE 
\STATE Let $i$ be the index s.t. $H_v[i]$ is the smallest $\tzheavy$ child of $v$ greater than or equal to $w$
\STATE Forward to a heavy node through port $P_v[i]$
\ENDIF
\end{algorithmic}
{\bf end operation}
\caption{The TZ scheme. Code for node $v$ for a message sent to node $w$.}
\label{algo: tz}
\end{algorithm}

\subsection{The CompactFTZ scheme}
\label{subsec: tzftdynamic}

\begin{figure}[t!]
\centering
\includegraphics[scale=0.50]{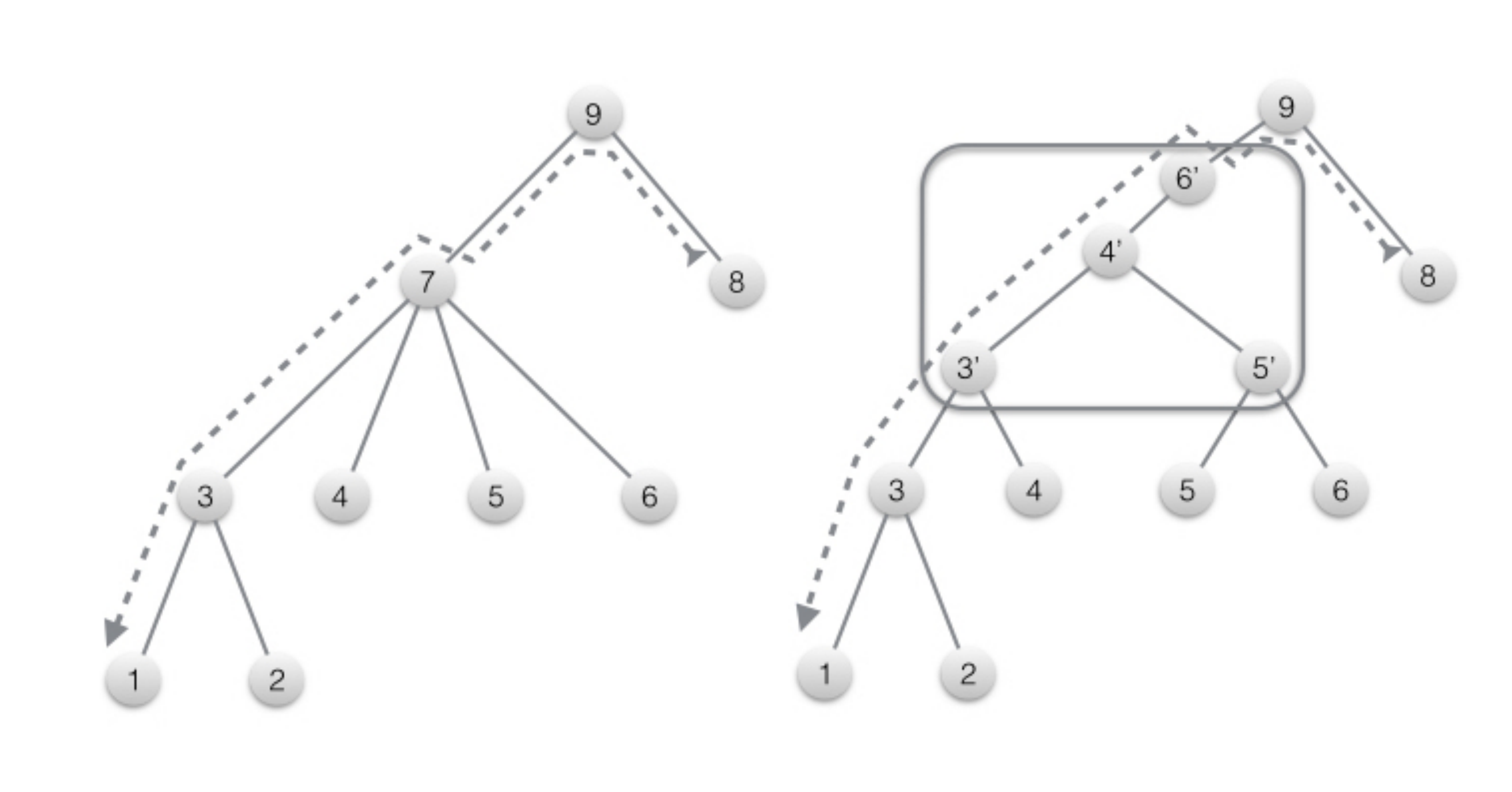}
\caption{ The left side shows the tree before any deletion with  the path a message from 8 to 1 will follow.
 The right side shows the tree obtained after deleting 7. The nodes enclosed in  the rectangle are virtual helper nodes replacing 7.
  To route a message from 8 to 1, virtual nodes perform binary search, while real nodes follow TZ.}
 \label{fig: example-routing}
\end{figure}

\begin{algorithm}[t!]

\textbf{Preprocessing:} \label{algoline: tzftinit}
Construct a BFS spanning tree of the network from an arbitrary node. Do a DFS labelling and $\tz$ setup followed by $\CFT$ data structures setup using $\tz$ DFS numbers as node $ID$s. 
\begin{algorithmic}[1]
\STATE $v$ runs $\CFT$ at all times.
\IF{$v$ is a real node}
\STATE Invoke $TZ_v(w, L(w))$ \label{01a}\\
\ELSE[$v$ is a virtual helper node (= $\helper(v))$]
\IF{$v = w$} \label{02a}
\STATE The message has reached its destination
\ELSIF{$w \notin [d_v, v]$} \label{04a}
\STATE Forward to the parent of $\helper(v)$ in the current virtual tree.   \label{05a}
\ELSIF{$w < v$}  \label{06a}
\STATE Forward to the left child of $\helper(v)$ in the current virtual tree.  \label{07a}
\ELSE  \label{08a}
\STATE Forward to the right child of $\helper(v)$ in the current virtual tree.  \label{09a}
\ENDIF
\ENDIF
\end{algorithmic}
\caption{ The $\tzft$ scheme. Code for node $v$ for a message sent to node $w$.}
\label{algo: compactftz}
\end{algorithm} 

 $\tzft$ (Algorithm~\ref{algo: compactftz}) combines $\tz$  with  $\CFT$  to make $\tz$ fault tolerant in the self-healing model.  The \emph{initialisation} phase (Algorithm~\ref{algo: compactftz} line \ref{algoline: tzftinit}) performed during preprocessing sets up the data structures for $\tzft$ in the following order: A BFS spanning tree of the network is constructed rooted at an arbitrary node, then a DFS labelling and $\tz$ setup is done as in Section~\ref{subsec: tzftstatic}, followed by $\CFT$ data structures setup using the previously generated DFS numbers as node $ID$s. The underlying layer is aware of the node $ID$s, DFS number $ID$s and node labels to be used for sending messages (as in $\tz$). 

Recall that, in our model, if there is no edge between $u$ and $v$,
and port numbers $x$ and $y$ of $u$ and $v$, respectively, are not in use,
then $u$ or $v$ can request an edge $(u,v)$ attached to these ports. 
In what follows, we assume that in $\CFT$, when a child $x$ of $p$ is deleted and a child $w$ of $x$
creates an edge $(p,w)$, such an edge will use the port of $p$ used by $(p,v)$ and any available port of~$w$.

Every node runs $\CFT$ at all time. For routing, a real node just follows $\tz$ (Algorithm~\ref{algo: compactftz}, Line~\ref{01a}),
while a virtual node first checks if the packet reached its destination (Line~\ref{02a}),
and if not, it performs a binary search over the current virtual tree
(Lines~\ref{04a} to~\ref{09a}). As mentioned earlier, though we use the notion of light and heavy nodes
in the initial setup and use it to compute routing tables and labels, 
we do not maintain this notion as the algorithm progresses but just use the initially assigned labels throughout. 
Further, following $\CFT$, if a node $x$ is deleted, it is replaced by $\RT(x)$. If a packet traverses $\RT(x)$,
 the virtual nodes ignore the heavy/light classification and just use the $ID$s to perform binary search.
Figure~\ref{fig: example-routing} illustrates $\tzft$ in action. In the figure, node $8$ sends a packet for node $1$. 
If there is no deletion in the tree, the packet will simply follow the path via the root $9$, node $7$, node $3$ to node $1$. Recall that at each node,
the node checks if the packet destination falls in the intervals given by its heavy node, otherwise, it uses the light routing index to pick the correct port to forward the message from the label of the destination node given in the message.
However, if node $7$ is deleted by the adversary, using $\CFT$, the children of $7$ construct $\RT(7)$ (recall this is also 
done in a compact manner). Since node $9$ has $\helper$ node $\helper(6)$ at the port where it had node $7$ earlier, the  packet gets forwarded to node $\helper(6)$.
Since $1$ is less than $6$, the packet traverses the left side of $\RT(7)$ and eventually reaches node $3$. Node $3$ applies the $\tz$ routing rules as before
 and the packet reaches node $1$.




We use the following notation: Let $T_t$ be the $\tzft$ tree after $t$ deletions.
For a vertex $v$,  let $T_t(v)$ denote the subtree of $T_t$ rooted at $v$.
The set with the children of $v$ in $T_t$ is denoted as 
$children_t(v)$, while $parent_t(v)$ is the parent of $v$ in $T_t$.
The set of $\ID$s in $T_t(v)$ is denoted as $\ID(T_t(v))$.
If $v$ has two children, $left_t(v)$ ($right_t(v)$) denotes the
left (right) child of $v$, and $L_t(v)$ ($R_t(v)$) denote
the left (right) subtree of $v$.
Given two nodes $u$ and $v$, 
we write $u < \ID(T_t(v))$ if $\ID(u)$ is smaller than any $\ID$ in $\ID(T_t(v))$,
and similarly, we write  $\ID(T_t(u)) < \ID(T_t(v))$ if every 
$\ID$ in $\ID(T_t(u))$ is smaller than any $\ID$ in $\ID(T_t(v))$.
The definitions naturally extends to $> , \leq$ and  $\geq$.


Theorem~\ref{theo-1} states the routing properties of $\tzft$. Lemma~\ref{lemma-bst} is the key to proving Theorem~\ref{theo-1} (proofs deferred to Section~\ref{sec: tzft-apdx}). Lemma~\ref{lemma-bst} basically states that, after any sequence of deletions and subsequent self-healing, real nodes maintain the $\tz$ properties and the helper nodes (i.e. the $\RT$s) the BST properties, allowing routing to always function.

\begin{lemma}
\label{lemma-bst}
At every time $t$, the $\tzft$ tree $T_t$ satisfies  the following two statements:
\beginsmall{enumerate}
\item For every real node $v \in T_t$, 
for every $c \in children_t(v)$, $v > ID(T_t(c))$,  
and for every $c, d \in children_t(v)$ with $c < d$,
$ID(T_t(c)) < ID(T_t(d))$.

\item For every virtual node $\helper(v) \in T_t$,
$v \geq ID(L_t(v))$ and $v < ID(R_t(v))$.

\endsmall{enumerate}

\end{lemma}

\begin{theorem}
\label{theo-1}
For every $T_t$, for every two real nodes $u, w \in T_t$,
$\tzft$ successfully delivers a message from $u$ to $w$ 
through a path in $T_t$ of size at most $\delta(u,w) + y(\log \Delta - 1)$,
where $\delta(u,w)$ is the distance between $u$ and $w$ in $T_0$
and $y \leq t$ is the number of non-leaf nodes deleted to
get $T_t$.
\end{theorem}

Lemma~\ref{tzft-mem} states the memory usage of $\tzft$ leading to the final correctness theorem (Theorem~\ref{tztt-final}).

\begin{lemma}
\label{tzft-mem}
$\tzft$ uses only $O(log^2 n)$ memory per node to route a packet.
\end{lemma}

\begin{proof}
$\CFT$ uses only $O(\log n)$ local memory (Theorem~\ref{thm: cftmain}). The local fields of a node for routing have at most a constant number ($O(b)$) fields which are node references ($\log n$) size, thus, using $O(\log n)$ memory. The label of a node (which is the 'address' on a packet) is, however,  of $O(\log^2 n)$ size (since there can be $O(\log n)$ light nodes on a source-target path) and therefore, a node needs $O(\log^2 n)$ bits to process such a packet.
\end{proof}

Ignoring congestion issues, Lemma~\ref{tzft-mem} implies that a node can store and route unto $x$ packets using $x. \log^2 n$ local memory. 
\begin{theorem}
\label{tztt-final}
$\tzft$ is a self-healing compact routing scheme.
\end{theorem}

\begin{proof}
The theorem follows directly from Lemma~\ref{tzft-mem}, Theorems~\ref{theo-1} and \ref{thm: cftmain}.
\end{proof}

\subsection{Reporting Non-delivery (deleted receivers and sources)}

Contrary to what happens in static schemes such as Thorup-Zwick~\cite{ThorupZ01}, 
we now have the issue that a node might want to send
a packet to a node that has been deleted in $G_t$,
hence we need a mechanism to report that a packet could not be delivered.
To achieve that, the header of a packet now is defined as follows:
when a node $s$ wants to send a packet to $t$,
it sends it with the header $((t, L(t) \cdot (s, L(s))$.
When running $\tzft$, each node considers only the first pair.

When a node $v$ receives a message $M$ with a header containing two pairs,
it proceeds as follows to detect an error i.e. a non-deliverable. The following conditions suggest to $v$ that the receiver $t$ has been deleted and the packet is non-deliverable:
\begin{enumerate}
\item \emph{If $v$ is a leaf (real node) and $v \ne t$:} This is a dead end since the packet cannot traverse further. This implies that $t$ must have been in the subtree of $v$ but the subtree of $v$ is now empty.
\item \emph{If $v$ is a non-leaf node but there is no node at the port it should forward to:} Similar to above, it indicates that $v$'s subtree involving $t$ is empty.
\item \emph{If $u$ sent the packet to $v$ but according to the routing rules, $v$ should send the packet back to $u$:} This happens when  $v$ is a helper node which is part of $\RT(t)$ or $\RT(x)$ where $x$ was not on the path $s-t$ in $T_0$. Node $v$ will receive the message either on the way up (towards the root) or way down (from the root). In either case, if $v$ is part of $\RT(t)$, due to the dfs numbering, it would have to return $M$ to $u$. Another possibility is that due to a number of deletions $\RT(t)$ has disappeared but then $x$ would either be an ascendent (if $M$ is on the way up) or $x$ would be a descendent of $t$ (if $M$ is on the way down). Either way, the DFS numbering would indicate to $v$ that it has to return the message to $u$. 
\end{enumerate}


If a target deletion has been detected due to the above rules,  $v$ removes the first pair of the header and sends back $M$ to 
the node it got $M$ from (with the header now only having $(s, L(s)$). When a node $v$ receives a message $M$ with a header containing only one pair,
it proceeds as before and applies the same rules discussed previously. This time, a non-delivery condition however implies that the source has been removed too, and, therefore $M$ can be discarded from the system. This ensures that `zombie' or undeliverable messages do not clog the system. 


\subsubsection{About stretch}

The stretch of a routing scheme $A$, denoted $\lambda(A,G)$, is the minimum $\lambda$ such that
$r(s,t) \leq \lambda \, dist(s,t)$ for every pair of nodes $s, t$, where 
$dist(s,t)$ is the distance between $s$ and $t$ in the graph $G$ and 
$r(s,t)$ is the length of the path in $G$ the scheme uses for routing a message
from $s$ to $t$. 

The stretch $\lambda(\tzft, T_0)$ is 1: for any pair of nodes, TZ routes a message through 
the unique path in the tree between them.
Similarly, the stretch $\lambda(\tzft, T_t)$ is 1:
each node that is deleted is replaced with a binary tree structure $R$,
and the nodes in it perform a binary search, hence a message passing through $R$ follows
the shortest path from the root to a leaf, or vice versa.

However, the stretch of $\tzft$ is different when we consider $G_t$.
First note that the stretch $\lambda(\tzft, G_0)$ might be of order $\Theta(n)$ since a spanning tree of a graph may blow up the distances by that much.
Since $\lambda(\tzft, T_0) = 1$, it follows that 
$\delta_T(u,w) \leq \lambda(\tzft, G_0) \cdot \delta_G(u,w)$,
where $\delta_T(u,w)$ is the distance between $u$ and $w$ in $T_0$ and
$\delta_G(u,w)$ is the distance between $u$ and $w$ in $G_0$.
Theorem~\ref{theo-1} states that, for routing a message from $u$ to $w$, 
$\tzft$ uses a path in $T_t$ of size at most $\delta_T(u,w) + y(\log \Delta - 1)$,
where  $y \leq t$ is the number of non-leaf nodes deleted to get $T_t$.
The $y(\log \Delta - 1)$ additive factor in the expression is because each deleted non-leaf node is replaced
with a binary tree, whose height is $O(\log \Delta)$.
In the worst case, that happens for all $y$ binary trees for a given message,
which implies that $\lambda(\tzft, G_t) \leq y(\log \Delta - 1) \cdot \lambda(\tzft, G_0)$ i.e. $\lambda(\tzft, G_t) \leq y(\log \Delta - 1) \cdot \lambda(\tzft, G_0)$ (since $\tzft$ only uses the tree for routing).

\section{CompactFT - Detailed Algorithms and Proofs}
\label{sec: cft-apdx}

%

\begin{algorithm}[th!]
\caption{CompactFT(Graph $G$): Main function}
\label{algo: cft}
\begin{algorithmic}[1]
\STATE \emph{Preprocessing and INIT:} 
A rooted BFS spanning tree $T(V,E')$ of $G(V,E)$ is computed. For every node $v$, its Will (non-leaf or leaf as appropriate) is computed.  Every node $x$ in a Will is labeled as $(x,<x>)$ where $x$ is $x$'s ID and $<x>$ is $x$'s port number at which $x$ is connected to its parent (if any). Each node only has a $\willportion$ and/or $\leafwillportions$ (constant sized (in number of node $ID$s) portion of parent's will or sibling's will respectively).\\
 Port number $0$ is reserved for  $v$'s parent in $T$. The children of $v$ are attached at ports $1$ to $\numchildren(v)$, where $\numchildren(v)$ is also the number of $v$'s children in $T$.
\WHILE {true}
\IF{a vertex $x$ is deleted}
\IF { $\numchildren(x) > 0$}
\STATE \textsc{FixNonLeafDeletion($x$)}
\ELSE 
\STATE \textsc{FixLeafDeletion($x$)}
\ENDIF
\ENDIF
\ENDWHILE
\end{algorithmic}
\end{algorithm}

%

\begin{algorithm}[h!]
\caption{\textsc{FixNonLeafDeletion($x$)}: Self-healing on deletion of internal node }
\label{algo: fixnode}
\begin{algorithmic}[1]
\WHILE{true}
\FOR{child $v$ of $x$ (if they exist)}
\STATE $\execute$ the Will of $x$ using the $O(\log n)$ sized $\willportion(d,v)$. \COMMENT{i.e. make the connections given by $\willportion(x,v)$}
\ENDFOR
\FOR{parent $t$ of $x$ (if it exists)}
\STATE $t$ will be contacted by heir of $x$, say $h$ to open a new connection to $h$ at port of deleted node, port $<x>$. 
\STATE $t$ will be informed by $h$ that the ID of the deleted node was $d$
\STATE Send $\BrNodeReplace((x,<x>),(h,<x>))$ message to every neighbour.\\
\ENDFOR
\IF{node $v$ receives message $\BrNodeReplace((x,<x>),(h,<x>))$ from parent $p$}
\STATE $v$ changes every occurrence of node $x$ to node $h$ in its $\willportion(x,v)$.
\ENDIF
\ENDWHILE
\end{algorithmic}
\end{algorithm}

 \begin{algorithm}[h!]
\caption{\textsc{FixLeafDeletion($x$)}: Self-healing on deletion of leaf node}
\label{algo: fixleaf}
\begin{algorithmic}[1]
\STATE Let $p=\parent(x)$
\IF[$p$ has not been deleted yet]{$p$ is a real node}
\STATE $p$ broadcasts $\BrLeafLost(<x>)$. \COMMENT{$p$ does not know $ID$ of $d$, only the port number $<d>$}
\IF{node $v$ receives a $\BrLeafLost(<x>)$ message}
\STATE $v$.\textsc{UpdateLeafWillPortion}$(p,<x>)$\COMMENT{Use $<x>$'s $\leafwill$ to update $p$'s will by updating $\willportions$ using broadcast($Br*$ messages) and/or point-to-point($Pt*$ messages)}
\ENDIF
\ELSE[The real parent of $x$ was already deleted earlier]
\IF[Note: Nodes $v$ and $d$ were neighbours]{node $v$ is $<x>$'s $\leafheir$}
\STATE  $\execute$ $<x>$'s $\leafwill$ \COMMENT{A $\leafwill$ has a $\bypass$ and/or a node takeover action}
\ENDIF
\IF[Note: Nodes $v$ and $d$ were neighbours]{node $x$ was $<v>$'s $\leafheir$}
\IF[$v$'s helper node is real $v$'s parent]{$\parent(v) = helper(v)$}
\STATE  $v$ designates the other child (i.e. not $v$) of $\parent(v)$ as $<v>$'s $\leafheir$ \COMMENT{Each node in the $\RT$has two children.}
\ELSE
\STATE  $v$ designates $\parent(v)$  as  new $\leafheir$ 
\ENDIF
\STATE $v$ sends $\leafwill(v)$ to $\leafheir(v)$
\ENDIF
\ENDIF
%
%
%
\end{algorithmic}
\end{algorithm}

 \begin{algorithm}[h!]
\caption{\textsc{UpdateLeafWillPortion($p,<x>$)}: Node $v$ updates Leaf Wills by `simulation'. The identity of any node $a$ is available as $(a,<a>)$ in the wills.}
\label{algo: updatewp}
\begin{algorithmic}[1]
\IF[Simulate execution of $x$'s $\leafwill$]{Node $v$ is $x$'s $\leafheir$}
\STATE Let $x'$ be the parent of $\helper(x)$ in $\will(p)$
 \IF[Case 1: $\helper(x)$ was not the parent of $\real(x)$ in $\will(p)$]{$\helper(v)$ is parent of $\real(x)$ in $\will(p)$}
\FOR{$\will(p)$}
\STATE Let $w$ be other child (i.e. not $x$) of $v$.
\STATE Let $u$ be the parent of $\helper(v)$.
\STATE Let $l'$ be the left child of $\helper(x)$
\STATE Let $r'$ be the right child of $\helper(x)$ \COMMENT{$NULL$ if $x$ was an heir.}
\ENDFOR
\STATE $p.\ptwillconnection((\real(w),<w>),(u,<u>))$ \COMMENT{Simulate $\bypass(<x>)$}
\STATE $p.\ptwillconnection((\helper(v),<v>),(x',<x'>))$ \COMMENT{$v$ sends messages through parent $p$}
\STATE $p.\ptwillconnection((\helper(v),<v>),(l',<l'>))$
\STATE $p.\ptwillconnection((\helper(v),<v>),(r',<r'>))$
\ELSE 
\STATE $p.\ptwillconnection((\helper(v),<v>),(x',<x'>))$ \COMMENT{Case 2: $\helper(x)$ was the parent of $\real(x)$; Only $\bypass(x)$ required.}
 \ENDIF
\ELSIF{node $x$ was $<v>$'s $\leafheir$}
\IF{$\parent(v) = \helper(v)$ in $\will(p)$}
\STATE  $v$ designates the other child (i.e. not $v$) of $\parent(v)$ as $<v>$'s $\leafheir$.
\ELSE
\STATE  $v$ designates $\parent(v)$  as  new $\leafheir$ 
\ENDIF
\STATE  $p.\ptnewleafwill((v,<v>),(\leafheir(v),<\leafheir(v)>),\leafwill(v))$ \COMMENT$v$ sends $\leafwill(v)$ to $\leafheir(v).$
\ENDIF
\end{algorithmic}
\end{algorithm}

\Armando{We should unify notation. I propose Amitabh sets it and I will follow.} 

\begin{lemma}
\label{lm: cftonehelper}
In $\CFT$, a real node simulates at most one helper node at a time. 
\end{lemma}
\begin{proof}
This follows from the construction of the algorithm. If  deleted node $x$ was a non-leaf node,  it is substituted by $\RT(x)$ (Figure~\ref{fig: RT}).
$\RT(x)$ is like a balanced binary tree such that the leaves are the real children of $x$ and each internal node is a virtual helper node. Also, there are exactly the same number of internal nodes as leaf nodes and each leaf node simulates exactly one helper node. This is the only time in the algorithm that a helper node is created. At other times, such as during leaf deletion or as a heir, a node may simulate a different node but this only happens if the node relinquishes its previous helper node. Thus, a real node simulates at most one helper node at any time.
\end{proof}

\begin{lemma}
\label{lm: deg}
The $\CFT$ increases the degree of any vertex by at most 3.
\end{lemma}
\begin{proof}
Since the degree of any node in a binary tree is at most 3, this lemma follows from Lemma~\ref{lm: cftonehelper}.
\end{proof}

\begin{lemma}
\label{lm: cftstructure}
The $\CFT$'s diameter is bounded by $O(D \log \Delta)$, where $D$ is the diameter  and $\Delta$ the highest degree of a node in the original graph ($G_0$).
\end{lemma}

\begin{proof}
This also follows from the construction of the algorithm. The initial spanning tree $T_0$ is a BFS spanning tree of $G_0$; thus, the diameter of $T_0$ may be at most 2 times that of $G_0$. Consider the deletion of a non-leaf node $x$ of degree $d$. $x$ is replaced by $\RT(x)$ (Figure~\ref{fig: RT}). Since $\RT(x)$ is a balanced binary tree (with an additional node), the largest distance in this tree is $\log d$. Two $\RT$s never merge, thus, this $\RT$ cannot grow. A leaf deletion can only reduce the number of nodes in a $\RT$ reducing distances. Consider a path which defined the diameter $D$ in $G_0$. In the worst case, every non-leaf node on this path was deleted and replaced by a $\RT$. Thus, this path can be of length at most $O(D \log \Delta)$ where $\Delta$ is the maximum possible value of $d$.
\end{proof}


\begin{lemma}
\label{lm: leafheirs}
In $\CFT$, a node may be $\leafheir$ for at most two leaf nodes.
\end{lemma}
\begin{proof}
For contradiction, consider a node $y$ that is  $\leafheir(u)$, $\leafheir(v)$ and $\leafheir(w)$ for three leaf nodes $u$, $v$ and $w$ all of whose parent is node $p$. From the construction of the algorithm, $v$'s $\leafheir$ can either be the parent of $\real(v)$ or a child of $\helper(v)$ in  $\RT(p)$. If node $y$ is $\real$, it cannot have a child in $\RT(p)$, therefore it can be $\leafheir$ for only one of $u$, $v$ or $w$ (by the first rule). However, if $y$ is a helper node, the following case may apply: $y$ is the parent of  both $\real(u)$ and $\real(v)$ and child of $\helper(w)$ (wlog): However, since the $\RT$ is a balanced binary search tree ordered on the $ID$s of the leaf children (the $\real$ nodes), one of $y$'s children (the left child) must be $\real(y)$. Therefore, $\helper(y)$ can be $\leafheir$ for either $u$ or $v$, and $w$.
\end{proof}

\begin{lemma}
\label{lm: cftmemory}
$\CFT$ requires only $O(\log n)$ memory per node.
\end{lemma}
\begin{proof}
Here, we analyse the memory requirements of a real node $v$ in $\CFT$. As mentioned before, $v$ does not store the list of its neighbours or its $\RT$ (these can be of $O(n)$ size). However, $v$ uses the $O(\log n)$ sized fields \fieldnumchildren\ and \fieldmaxportnum\ to keep track of the number of children it presently has and the maximum port number it uses. 
$\CFT$ uses the property that the initial port assignment does not change. 
If $v$ is a non-leaf node, it does not store any piece of $\will(v)$ that is distributed among $v$'s children. If $v$ is a leaf node, it has only a $O(\log n)$ sized will ($\leafwill(v)$), which it stores with its $\leafheir$ (though $v$ can also store the $\leafwill(v)$).  Let $p$ be the parent of $v$. If $p$ is $\real$, Node $v$ will store $\willportion(p,v)$, \ie the portion of $\will(p)$ (\ie $\RT(p)$) in which $v$ is involved. From Lemma~\ref{lm: cftonehelper}, there can only be two occurrences of $v$ (one as $\real$ and one as $\helper$). Since $\RT(p)$ is a binary tree, $\helper(v)$ can have at most 3 neighbours and $\real(v)$ at most 1 (as real nodes are leaves in a $\RT$). Therefore, the total number of $\ID$s in $\leafwillportion(p,v)$ cannot exceed 6 and its size is $O(\log n)$. By the same logic, if $v$ hosts a $\helper$ node, by Lemma~\ref{lm: cftonehelper}, it only requires $O(\log n)$ memory. $v$ may also have $\leafwill$s for its siblings. By Lemma~\ref{lm: leafheirs}, $\real(v)$ and $\helper(v)$ may store at most 2 of these wills each; this takes only $O(\log n)$ storage. Every node in a $\will$ is identified by both its $ID$ and port number. However, this only doubles the memory requirement. Finally, all the messages exchanged (Table~\ref{tab: CFTmessages}) are of $O(\log n)$ size.
\end{proof}

\othm{thm: cftmain}
The $\CFT$ has the following properties:
\begin{enumerate}
\item \label{thm: cftdeg} $\CFT$ increases the degree of any vertex by only 3.
\item \label{thm: cftdiam} $\CFT$ always has diameter bounded by $O(D \log \Delta)$, where $D$ is the diameter  and $\Delta$ the maximum degree of the initial graph.
\item \label{thm: cftmem} Each node in $\CFT$ uses only $O(\log n)$ local memory  for the algorithm.
\item \label{thm: cftcosts} The latency per deletion is $O(1)$ and the number of messages sent per node per deletion is $O(\Delta)$; each message contains $O(1)$ node $\ID$s and thus $O(\log n)$ bits.
\end{enumerate}
\eothm

\Armando{We can say something stronger about the diameter, right? The diameter increases only when
a non-leaf node is deleted, so after $t$ deletion, the diameter is at most $O(D + y \log \Delta)$,
where $y \leq t$ is the number of non-leaf deletions.}
\Amitabh{Ooops. I think diameter is at most $O(D \log \Delta)$ is a stronger claim than $O(D + x \log \Delta)$, imagine D as constant and x as $O(n)$}

\begin{proof}
Part~\ref{thm: cftdeg} follows from Lemma~\ref{lm: deg} and Part~\ref{thm: cftdiam} from Lemma~\ref{lm: cftstructure}. Part~\ref{thm: cftmem} follows from Lemma~\ref{lm: cftmemory}. Part~\ref{thm: cftcosts} follows from the construction of the algorithm. Since the virtual helper nodes have degree at most 3, healing one deletion results in at most O(1) changes to the edges in each affected reconstruction tree. As argued in Lemma~\ref{lm: cftmemory}, both the memory and any messages thus constructed are $O(\log n)$ bits. Any message is required to be sent only $O(1)$ hops away. Moreover, all changes can be made in parallel. Only the broadcast messages, $Br*$ messages, are broadcasted by a $\real$ node to all its neighbours and thus $O(\Delta)$ messages (sent in parallel) may be used.
\end{proof}

\section{CompactFTZ  - Proofs}

\label{sec: tzft-apdx}

\olem{lemma-bst}
At every time $t$, the $\tzft$ tree $T_t$ satisfies  the following two statements:

\begin{enumerate}

\item For every real node $v \in T_t$, 
for every $c \in children_t(v)$, $v > ID(T_t(d))$,  
and for every $c, d \in children_t(v)$ with $c < d$,
$ID(T_t(c)) < ID(T_t(d))$.

\item For every virtual node $v' \in T_t$,
$v \geq ID(L_t(v))$ and $v < ID(R_t(v))$.

\end{enumerate}

\eolem

\begin{figure}[ht!]
\centering
\includegraphics[scale=0.4]{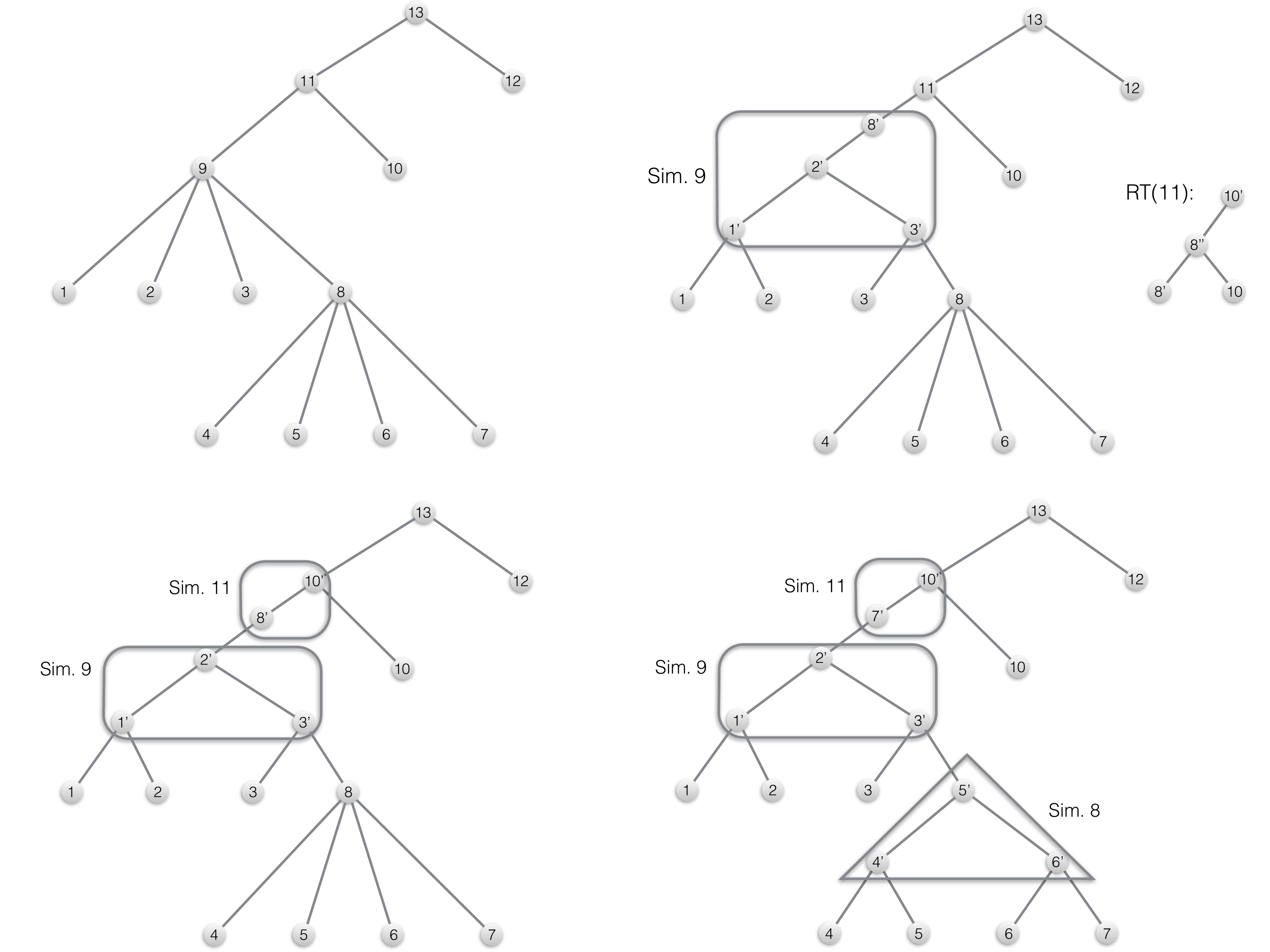}
\caption{A sequence of deletions.}
 \label{fig: example-proof}
\end{figure}

\begin{proof}

We proceed by induction on $t$.
For $t=0$, $T_0$ satisfies property (1) because
its the properties of initial DFS labelling, while
property (2) is satisfied since there are no virtual nodes in $T_0$.

Suppose that $T_t$ satisfies (1) and (2).
Let $v \in T_t$ the node deleted to obtain $T_{t+1}$.
We show $T_{t+1}$ satisfies (1) and (2), we only need to
check the nodes that are affected when getting $T_{t+1}$.
We identify two cases:

\begin{enumerate}

\item \underline{$v$ is not a leaf in $T_t$}.
Let $c_1, \hdots, c_x$ be the children of $v$ in $T_t$ in ascending order.
Each $c_i$ might be real or virtual, and if it is virtual then 
it is denoted $c'_i$ and is simulated by a real node $real(c'_i)$.
Let $RT(v)$ be the reconstruction tree obtained from the children
of $v$, as explained in Section~\ref{sec: cft}.
By construction, we have that (virtual) $c'_x$, the child of $v$ with the largest ID, is the root
of $RT(v)$ and it has no right subtree. Moreover, $c'_{x-1}$ is the left child of $c'_x$.
We now see what happens in $T_{t+1}$.
We first analyze the case of the parent of $v$ 
and then the case of the children.

If $z=parent_t(v)$ is a real node, then $z$ and 
$c'_x$ create an edge between them, and hence $RT(v)$ is a subtree of $z$ in $T_{t+1}$
(see Figures~\ref{fig: example-proof} top-left and top-right where the node 9 is deleted).
By induction hypothesis, the lemma holds for $v$ in $T_t$ and $RT(v)$ contains only 
the children of $v$, hence the lemma still holds for $z$ in $T_{t+1}$.
Now, if $z'=parent_t(v)$ is a virtual node, 
then it must be that there is a virtual node $v'$ in $T_t$ that
is an ancestor of $v$ (this happens because at some point
the parent of $v$ in $T= T_0$ was deleted, hence $v$ created a virtual node $v'$;
see Figure~\ref{fig: example-proof} bottom-left where the parent of 8 is virtual).
For now, suppose $c_x$ is a real node (below we deal with the other case).
In $T_{t+1}$, $c'_x$ replaces $v'$,
and $z'$ and $c'_{x-1}$ create an edge
between them, hence in the end
$left_{t+1}(c'_x) = left_t(v')$, $right_{t+1}(c'_x) = right_t(v')$ and 
$z' = parent_{t+1}(c'_{x-1})$ (see Figure~\ref{fig: example-proof} bottom-right where 7' replaces 8').
In other words, the root $c'_x$ of $RT(v)$ takes over $v'$
and the left subtree of $c'_x$ in $RT(v)$ (whose root is $c'_{x-1}$) is connected to $z'$.
We argue that the lemma holds for $c'_x$ and $z'$ in $T_{t+1}$.
The case of $z'$ is simple: by induction hypothesis, the lemma holds for 
$z'$ and $v$ in $T_t$, and $RT(v)$ is made of the children of $v$ in $T_t$.
The case of $c'_x$ is a bit more tricky.
First, since the lemma holds for $v'$ in $T_t$, then
it must be that $v \in L_t(v')$ and $c_x < ID(R_t(v'))$.
Also, we have that $v < c'_x$, hence 
$c'_x \geq ID(L_{t+1}(c'_x))$ and
$c'_x < ID(R_{t+1}(c'_x))$.

Now, let's see what happens with a child $c_i$.
If $c_i$ is a real node, then virtual $c'_i$ (which belongs to $RT(v)$)
is an ancestor of $c_i$ in
$T_{t+1}$. The lemma holds for $c_i$ because the subtrees of
$c_i$ in $T_t$ and $T_{t+1}$ are the same. 
Similarly, the lemma holds for $c'_i$ because $RT(v)$
is a binary search tree and the lemma holds for each child of $v$
in $T_t$, by induction hypothesis.

Consider now the case $c_i$ is a virtual node, hence denoted $c'_i$.
In this case, in $T_t$, (real) $c_i$ is a descendant of $c'_i$.
We have that $c'_i$ appears in $RT(v)$ two times, and both of them as a virtual node,
one as a leaf and the other as an internal node
(see Figure~\ref{fig: example-proof} top-right where 11 is deleted to get the tree at the 
bottom-left; $8'$ is virtual, is a child of $11$ and appears two times as virtual node in $RT(11)$).
Let $c'_i$ denote the virtual leaf (this node also belongs to $T_t$)
and $c''_i$ denote the internal virtual node.
So, by replacing $v$ with $RT(v)$, it would not be true any more
that every real node simulates at most one virtual nodes,
since $c_i$ would be simulating $c'_i$ and $c''_i$.
To solve this situation, $u' = left_t(c'_i)$ replaces $c'_i$,
and $c'_i$ replaces $c''_i$ (in this case $c'_i$ always has only one 
child in $T_t$, $u'$, which is left).
In other words, in $R(v)$, $c'_i$ is moved up to the position of $c''_i$
and $u'$ is moved up to the position of $c'_i$
(see Figure~\ref{fig: example-proof} bottom-left).
Thus, $L_{t+1}(u') = L_t(u')$
and $R_{t+1}(u') = R_t(u')$.
The lemma holds for $u'$ because it was moved up one step
and the lemma holds for it in $T_t$, by induction hypothesis.
To prove that the lemma holds for $c'_i$, let us consider 
$RT(v)$ and $c'_i, c''_i$ in it. By construction, we have that 
$c'_i \geq ID(L(c''_i, RT(v)))$ and 
$c'_i < ID(R(c''_i, RT(v)))$.
Also, since the lemma holds for each child of $v$ in $T_t$,
for each child $c_j \neq c'_i$ of $v$ in $T_j$, if $c_j < c'_i$,
it must be that $c'_i > ID(T_t(c_j))$, otherwise $c'_i < ID(T_t(c_j))$.
These two observation imply that 
$c'_i \geq ID(L_{t+1}(c'_i))$ and $c'_i < ID(R_{t+1}(c'_i))$.
This completes the case.

\item \underline{$v$ is a leaf in $T_t$}.
First consider the subcase when the $parent_t(v)$ is a real node.
We have that $T_{t+1}$ is $T_t$ minus the leaf $v$, hence the lemma holds for $T_{t+1}$,
by induction hypothesis.

Now, if $u' = parent_t(v)$ is a virtual node then in $T_t$, there is a virtual node 
$v'$, which is an ancestor of $v$ (this happens because at some point the paren
of $v$ in $T_0 = T$ was deleted, hence $v$ created a virtual node).
In $T_t$, there is descending path from $v'$ to $v$, that passes through 
virtual nodes until reaches $v$. Let us denote this path
$P = v', u'_1, \hdots, u'_x, v$, for some $x \geq 0$.
We have the following three subcases.
 
If $x=0$, then $v'$ is the parent of $v$, and actually $left_t(v') = v$,
by the induction hypothesis. Thus, $v'$ and $v$ are just removed, 
and $R_t(v')$ replaces $v'$ in $T_{t+1}$, namely,
$parent_t(v')$ is connected the root of $R_t(v')$.
It is easy to check that the lemma holds for $T_{t+1}$.

If $x=1$ then 
$u'_1$ replaces $v'$ and $L_{t+1}(u'_1) = L_t(u'_1)$
and $R_{t+1}(u'_1) = R_t(v')$.
Namely, $v$ and $v'$ are removed and $u'_1$ is moved up one step.
Again, it is not hard to see that the lemma holds for $T_{t+1}$.

The last subcase is $x > 1$. In $T_{t+1}$, $u'_x$ replaces $v'$ and 
$R_{t+1}(u'_{x-1}) = L_t(u'_x)$.
Clearly, the lemma holds for $u'_{x-1}$ because $u'_{x-1} < ID(L_t(u'_x))$,
by induction hypothesis.
To prove that the lemma holds for $u'_x$, we observe the following about
the path $P = v', u'_1, \hdots, u'_x, v$, $x > 1$
(e.g., the path from 7' to 7 in Figure~\ref{fig: example-proof} bottom-right).
The lemma holds for $T_t$,
hence $v \in L_t(v')$, which implies that $u'_1 = left_(v')$,
and actually $u'_i < v' = v$, for each $v_i$.
Also observe that the induction hypothesis implies that 
$v \in R_t(u_i)$, for each $u'_i$.
Therefore, we have that  $u'_x > ID(T_t(u'_i))$, $1 \leq i \leq x-1$,
which implies that $u'_x > ID(L_{t+1}(u'_x))$.
Also, since $u'_x \in L_t(v')$, it must be that $u'_x < ID(R_t(v'))$,
hence $u'_x < ID(R_{t+1}(u'_x))$. The induction step follows.
\end{enumerate}
\end{proof}


\othm{theo-1}
For every $T_t$, for every two real nodes $u, w \in T_t$,
$\tzft$ successfully delivers a message from $u$ to $w$ 
through a path in $T_t$ of size at most $\delta(u,w) + y(\log \Delta - 1)$,
where $\delta(u,w)$ is the distance between $u$ and $w$ in $T_0$
and $y \leq t$ is the number of non-leaf nodes deleted to
get $T_t$.
\eothm

\begin{proof}
The claim holds for $t=0$ (static case, analogous to~\cite{ThorupZ01}).
For $t > 0$, suppose a message $M$ from $u$ to $w$ in $T_t$, reaches a real node $x$
(possibly $x=u$). Note that $x$ is oblivious to the $t$-th node deletion, namely,
it does not change its routing tables in the healing-process, hence it makes the same decision
as in $T_{t-1}$. Let $\#portnumber$ be the port through which $x$ sends
$M$ in $T_t$, and let $v$ be the vertex that is connected to $x$ through port
$\#portnumber$ in $T_{t-1}$. 
Lemma~\ref{lemma-bst} (1) implies that if $w$ is ancestor (descendant) of 
$x$ in $T_{t-1}$, then it is ancestor (descendant) of $x$ in $T_t$.
Moreover, the path in $T_t$ from $x$ to $w$ passes through port $\#portnumber$.
If $v$ is not the $t$-th vertex deleted, then, $v$ is in $T_t$,
and it gets $M$ from $x$, which implies that $M$ gets closer to $w$. 
Otherwise, $v$ is the $t$-th vertex deleted. In this case, in $T_t$, a virtual 
node $y'$ is connected to $x$ through port $\#portnumber$,
thus, $y'$ gets $M$ from $x$.
By Lemma~\ref{lemma-bst} (2), $y'$ necessarily sends $M$ to a virtual or real node that is closer to $w$. 
Thus, we conclude that $M$ reaches $w$.

For the length of the path, note that after $t$ deletions, from which $y$ are non-leafs,
at most $y$ nodes in the path from $u$ to $w$ in $T_0$ are replaced in $T_t$
with $y$ binary trees of depth $O(\log \Delta)$ each of them.
Then, the length of the path from $u$ to $w$ in $T_t$ is at most
$\delta(u,w) + y(\log \Delta ) - y = \delta(u,w) + y(\log \Delta - 1)$, in case the path has
to pass through all these trees from the root to a leaf, or vice versa.
\end{proof}

\section{Extensions and Conclusion}
This paper presented, to our knowledge, the first compact self-healing algorithm and also the first self-healing compact routing scheme. We have not considered the memory costs involved in the preprocessing but we believe that it should be possible to set up the data structures in a distributed compact manner: this needs to be investigated.
The current paper focuses only on node deletions,.  Can we devise a self-healing compact routing scheme working in a fully dynamic scenario with both (node and edge) insertions and deletions? The challenges reside in dealing with the expanding out-degree efficiently.  

The current paper allows to add additional links to nearby nodes in an overlay manner.  What should the model be of losing links without losing nodes? How will it affect the algorithms appearing in this paper?



\pagebreak

{\small
\bibliographystyle{plain}
\bibliography{selfheal-routing}
}
\end{document}